\documentclass[12pt,reqno]{amsart}
\usepackage{amsmath}
\usepackage{amssymb}
\usepackage{amsthm}
\usepackage{mathrsfs}
\usepackage{latexsym}
\usepackage{xcolor}
\usepackage{multirow} 
\usepackage{comment}
\usepackage{booktabs}
\usepackage{graphicx}
\usepackage{subcaption}
\usepackage{float}
\usepackage{placeins}
\usepackage{hyperref}
\makeatletter
\renewcommand{\subsection}{%
  \@startsection{subsection}{2}%
  {\z@}%
  {1.2\baselineskip}%
  {0.7\baselineskip}%
  {\normalfont\bfseries}%
}
\makeatother
\newtheorem{theorem}{Theorem}[section]
\newtheorem{lemma}{Lemma}[section]
\newtheorem{corollary}{Corollary}[section]
\newtheorem{remark}{Remark}[section]
\newtheorem{definition}{Definition}[section]
\newtheorem{proposition}{Proposition}[section]
\newtheorem{example}{Example}[section]
\newtheorem{assumption}{Assumption}[section]
\numberwithin{equation}{section}
\newcommand{\bth}{\begin{theorem}}
\newcommand{\ethe}{\end{theorem}}

\newcommand{\bre}{\begin{remark}}
\newcommand{\ere}{\end{remark}}

\newcommand{\ble}{\begin{lemma}}
\newcommand{\ele}{\end{lemma}}
\newcommand{\bde}{\begin{definition}}
\newcommand{\ede}{\end{definition}}

\newcommand{\bco}{\begin{corollary}}
\newcommand{\eco}{\end{corollary}}

\newcommand{\bpr}{\begin{proposition}}
\newcommand{\epr}{\end{proposition}}

\newcommand{\bexer}{\begin{exercise}}
\newcommand{\eexer}{\end{exercise}}
\newcommand{\breh}{\begin{hint}}
\newcommand{\ereh}{\end{hint}}

\newcommand{\halmos}{\hfill \qed}

\newcommand{\bexam}{\begin{example}}
\newcommand{\eexam}{\end{example}}

\newcommand{\pr} {{\bf Proof.}}

\newcommand{\bfi}{\begin{fig}}
\newcommand{\efi}{\end{fig}}

\newcommand{\beao}{\begin{eqnarray*}}
\newcommand{\eeao}{\end{eqnarray*}\noindent}

\newcommand{\beam}{\begin{eqnarray}}
\newcommand{\eeam}{\end{eqnarray}\noindent}
\newcommand{\E}{\mathbf{E}}
\newcommand{\PP}{\mathbf{P}}
\newcommand{\e}{\mathrm{e}}
\newcommand{\dd}{\mathrm{d}}

\newcommand{\xto}{x\to\infty}

\newcommand{\bH}{\overline{H}}
\newcommand{\bF}{\overline{F}}

\newcommand{\bG}{\overline{G}}
\newcommand{\bV}{\overline{V}}
\newcommand{\bU}{\overline{U}}

\newcommand{\bbr}{{\mathbb R}}

\newcommand{\bbb}{{\mathbb B}}

\newcommand{\bbn}{{\mathbb N}}

\newcommand{\yto}{y\to\infty}

\newcommand{\vep}{\varepsilon}

\allowdisplaybreaks[1]

\begin{document}
\title[Continuous-time risk model with multivariate strongly regular variation]
%{Asymptotics for Continuous-time risk models with multivariate strongly regularly varying insurance and financial risks}
{Asymptotics for a Continuous-Time Risk Model with Multivariate Strongly Regularly Varying Claims and Stochastic Investment Returns}

\author[Z. Chen, C.D. Passalidis]{Zhangting Chen, Charalampos  D. Passalidis} 

\address{ School of Mathematical Sciences, Soochow University, Suzhou, Jiangsu, China}
\address{Dept. of Statistics and Actuarial-Financial Mathematics,
	University of the Aegean, Karlovassi, GR-83 200 Samos, Greece}

\email{20234007012@stu.suda.edu.cn} \email{sasd24009@sas.aegean.gr.}

\date{{\small \today}}

\begin{abstract}
In this paper we study the finite-time asymptotic behavior of the entrance probability of the discounted aggregate claims into an extreme region in a multidimensional risk model. Our non-standard risk model is not restricted to the L\'{e}vy-Renewal framework, that represents a usual assumption for the stochastic returns and the counting process of the claim vectors, respectively. For the modeling of the distribution of claim vectors, we introduce the notion of set-based multivariate (strongly) regularly varying distributions. For the new set-based classes we provide certain closure properties and some examples to show the difference from the usual definitions of multivariate regularly varying distributions. This set-based framework provides a broad and flexible framework to the actuarial applications, and facilitates the modeling of dependence among the claim vectors that arrive at different time epochs. The two main results of the paper establish a unified asymptotic expression, that demonstrates the presence of multivariate linear single big jump principle. Despite of the assumptions the asymptotic expression remains explicit. In the first main result we suppose that the common distribution of the claim vectors belongs to the class of set-based multivariate regularly varying distributions, and the stochastic returns satisfy a general moment condition. In the second main result, we restrict the claim vectors in the class of set-based multivariate strongly regularly varying distributions, relaxing the moment condition on the stochastic returns. Finally, we conduct a numerical study to illustrate the accuracy of the proposed asymptotic approximation and investigate its sensitivity to the main model parameters.
\end{abstract}

\maketitle
%\vspace{3mm}
\textit{Keywords: Multivariate risk model; Stochastic investment returns; Interdependence; Non L\'{e}vy-Renewal environment; Set-based multivariate (strongly) regular variation}
\vspace{3mm}

\textit{Mathematics Subject Classification}: Primary 62P05 ;\quad Secondary 60G70.

%\textit{Journal of Economic Literature}: Primary C46 ;\quad Secondary G32.

%\vspace{3mm}
 %\begin{center}
%\textbf{ Dedicated to the memory of \\{Alexander Alekseevich Borovkov (1931 - 2026)}}
% \end{center}

\section{Introduction} \label{sec.CKP28T.1}

\subsection{Model description} \label{subsec.CKP28T.1.1}

Modern insurance companies, in order to remain viable and competitive, operate multiple lines of business and invest their surplus in both risk-free and risky financial assets. Thus, the study of multivariate risk models in the presence of investment is of great importance for a more accurate assessment of insurance risks.

In this paper, we consider an insurer operating $d-$lines of business, where $d\in\bbn$, and investing its surplus in both risk-free and risky assets. More specifically, let $\{{\bf X}^{(i)},\,i\in\bbn\}$ denote the claim vectors representing the insurance risks, which arrive at times $\{\tau_i,\,i\in\bbn\}$, with $\tau_0=0$. The corresponding counting process $\{N(t),\,t\geq0\}$ is defined by
\beao
N(t)=\sup\{n\in\bbn:\tau_n\leq t\},
\eeao
with the convention $\sup\emptyset=0$. Let
\(
\lambda(t):=\E[N(t)]=\sum_{i=1}^{\infty}\PP(\tau_i\leq t),
\)
and assume that $\lambda(t)<\infty$ for every fixed $t\geq0$. We further define
\(
\Lambda:=\{t\geq0:\lambda(t)\in(0,\infty)\}.
\)

To avoid trivial cases, we consider that each claim vector ${\bf X}^{(i)}=(X_1^{(i)},\,\ldots,\,X_d^{(i)})^{\top}$ (where ${\bf y}^{\top}$ is the transpose of vector ${\bf y}$) may have zero components, but not all components are zero. Additionally, throughout this paper, the vectors $\{{\bf X}^{(i)},\,i\in\bbn\}$ are assumed to be identically distributed as a generic random vector ${\bf X}$ with common distribution $F$ supported on $\bbr_+^d:=[0,\infty)^d$. We further assume that the logarithmic return of the insurer's investment portfolio is described by a process $\{\xi(t),\,t\geq0\}$ with c\'{a}dl\'{a}g sample paths and $\xi(0)=0$.

Then, the discounted aggregate claims for the insurer up to a fixed time $T \in \Lambda$ are given by the relation
\beam \label{eq.CKP.1.1} 
{\bf D}(T) &:=& \sum_{i=1}^{\infty} {\bf X}^{(i)}\,\e^{-\xi(\tau_i)}\,{\bf 1}_{\{\tau_i \leq T\}}=\begin{pmatrix}   
\sum_{i=1}^{\infty} X_{1}^{(i)}\,\e ^{-\xi(\tau_i)}\,{\bf 1}_{\{\tau_i \leq T\}} \\ \vdots \\ \sum_{i=1}^{\infty} X_d^{(i)}\,\e ^{-\xi(\tau_i)}\,{\bf 1}_{\{\tau_i \leq T\}}  
\end{pmatrix}\,,
\eeam
where ${\bf 1}_E$ denotes the indicator of the event $E$.
Note that, for any fixed $T \notin \Lambda$, it holds ${\bf D}(T) = {\bf 0}$. For the sake of convenience in writing, we denote:
\[
\Theta_i(T) := \e^{-\xi(\tau_i)}\,{\bf 1}_{\{\tau_i \leq T\}}\,,\quad i \in \bbn\,.
\]

Our main objective is to study the asymptotic behavior, as $\xto$, of the probabilities
\beam \label{eq.CKP.1.4}
\PP\left({\bf D}(T)\in x\,A\right)\,,
%\qquad \PP\left({\bf D}(\infty)\in x\,A\right)\,,
\eeam
for various sets $A$ belonging to the general class $\mathscr{R}$, see Subsection \ref{subsec.CKP28T.2.2} below, with particular emphasis on their actuarial applications.

\subsection{Brief review and main contribution}
\label{subsec.CKP28T.1.2}

The asymptotic estimation of probabilities such as
\eqref{eq.CKP.1.4} has received considerable attention in the modern
risk theory literature, especially for heavy-tailed insurance risks. Under
suitable dependence conditions among the claim vectors, and moment conditions on the investment returns and on the counting process, the multivariate linear
single big jump principle typically yields an asymptotic relation of the
form
\beam \label{eq.CKP.1.5}
\lim_{x\to\infty}\frac{\PP\left({\bf D}(T)\in xA\right)}{\sum_{i=1}^{\infty}
\PP\left({\bf X}^{(i)}\Theta_i(T)\in xA\right)}
=1.
\eeam

Although \eqref{eq.CKP.1.5} captures the single big jump principle for
${\bf D}(T)$, its denominator still involves the tail probabilities
of randomly weighted claim vectors and may therefore be difficult to
evaluate explicitly. Under additional assumptions on the claim
distribution, the counting process $\{N(t),\,t\geq0\}$, and the
investment return process $\{\xi(t),\,t\geq0\}$, a more explicit
asymptotic formula can be obtained.

More concretely, in the one-dimensional case with $A=(1,\infty)$,
\cite{tang:wang:yuen:2010} proved that
\beao
\lim_{x\to\infty}
\frac{\PP(D(t)>x)}{\PP(X>x)}
=\int_0^t \e^{s\phi(\alpha)}\,\lambda(\dd s),
\eeao
uniformly for $t\in\Lambda\cup\{\infty\}$, provided that the claims are independent and identically distributed (i.i.d.) with a
regularly varying distribution of index $\alpha>0$, $\{N(t),\,t\geq0 \}$ is a renewal counting process, and that
$\{\xi(t),\,t\geq0\}$ is a L\'evy process with Laplace exponent
$\phi(\cdot)$ satisfying $\phi(p)<0$ for some $p>\alpha$. The claim
sequence, the counting process, and the investment return process were
also assumed to be mutually independent. Later, \cite{li:2012}
extended this result to a time-dependent risk model under a general
weak dependence structure between the claim sizes and inter-arrival
times. We also refer the reader to \cite{heyde:wang:2009,guo:2022}, for related contributions
in the one-dimensional setting. The literature on the one-dimensional
case is extensive, and we do not attempt to provide a comprehensive review here. However, our main results are also new in the one-dimensional setting.

In the multivariate setting, classical multivariate regular variation ($\operatorname{MRV}$, also used
for ``multivariate regularly varying'' throughout this paper)
provides a natural framework for describing joint tail behavior and has
been widely used in the asymptotic analysis of ruin probabilities for
multidimensional risk models; see, for example
\cite{konstantinides:li:2016,li:2016,
cheng:konstantinides:wang:2022,yang:su:2023}. A distinctive feature of this framework
is its global nature, since it describes the asymptotic behavior of the
entire multivariate tail measure. In many risk problems, however, the
quantity of interest is associated with a particular extreme region. For
example, write $D_k(T):=\sum_{i=1}^{\infty} X_k^{(i)}\,\e ^{-\xi(\tau_i)}\,{\bf 1}_{\{\tau_i \leq T\}}$ for $1\leq k\leq d$. If further we denote
\beam  \label{eq.CKP.1.4b}
A_{1} =\left\{{\bf x}\in\bbr^d:
\max_{1\leq k\leq d}x_k>1\right\}\in\mathscr{R},
\eeam
then
\[
\PP({\bf D}(T)\in x\, A_{1} )=
\PP\left(\max_{1\leq k\leq d}D_k(T)>x\right),
\]
which corresponds to the event that at least one component exceeds the
level $x$.

Another important set of family $\mathscr{R}$, with both theoretical and practical interest, is the set
\beam  \label{eq.CKP.1.4a}
A_2 = \left\{{\bf x}\;:\;\sum_{k=1}^d l_k\,x_k > b \right\}\,,
\eeam
with $l_k > 0$, for $k=1,\,\ldots,\,d$, $\sum_{k=1}^d l_k =1$ and $b> 0$. Thus, 
\[
\PP({\bf D}(T)\in xA_2)=
\PP\left(\sum_{k=1}^{d} l_k D_k(T)>bx\right),
\]
which describes the probability that the sum of discounted aggregate claims of the $d$-lines of business to exceed some high level $bx$ .

A set-based(in what follows, when a particular set is fixed, we replace
``set'' in ``set-based'' with the symbol denoting that set) approach to multivariate subexponential distributions was first developed by \cite{samorodnitsky:sun:2016}, and based on their approach \cite{konstantinides:passalidis:2024g}, \cite{konstantinides:passalidis:2025h} introduced many other multivariate heavy-tailed classes of distributions. \cite{samorodnitsky:sun:2016} characterize  the tail behavior
associated with a  set $A$ through
$\PP({\bf X}\in xA)$. Motivated by this point of view, we formulate the
tail behavior of the claim vector with respect to a fixed
$A\in\mathscr R$. This formulation also allows us to accommodate
dependence among different claim vectors. More precisely, instead of
assuming independence of $\{{\bf X}^{(i)},\,i\in\bbn\}$, we impose a
set-based tail quasi-asymptotic independence condition, see for the definition  Assumption~\ref{ass.CKP.3.1}.

Our goal in Theorem \ref{th.CKP.3.1a} is to extend the classical asymptotic estimations on the L\'{e}vy-Renewal environment, to a (multivariate) non-standard risk model, with interdependent claim vectors. To achieve this, we introduced the class of set-based $\operatorname{MRV}$ distributions, and we use two general moment assumptions on the stochastic discount factors and counting process, see Assumption \ref{ass.CKP.3.2a} and \ref{ass.CKP.3.3}.
Furthermore, we impose the set-based multivariate strongly regular variation ($\operatorname{MRV^\ast}$, also used for ``multivariate strongly regularly varying'' throughout this paper). This distributional choice for claim vectors is motivated by the relaxation of the moment condition \eqref{eq.CKP.3.3a} on the stochastic
investment return process. Most existing papers in the literature, use such  classical moment conditions through a finite condition on the Laplace exponent of the L\'{e}vy process $\xi$, see for example \cite{tang:wang:yuen:2010, yang:zhang:jiang:cheng:2015,li:2016,yang:wang:liu:zhang:2019}, among many others. In contrast to Theorem \ref{th.CKP.3.1a}, in Theorem \ref{th.CKP.3.1}, the strongly regular variation, enables us to
assume a tail-negligibility condition on the stochastic discount
factors generated by $\xi$, without requiring a moment of order strictly larger than the tail index of the distribution of claim vectors (see, Assumption \ref{ass.CKP.3.2} below).

To conclude, our main results show that, under a weak (but very general) dependence condition on
the claim vectors and suitable mild assumptions on the counting and investment processes, it holds that
\beam\label{eq.CPK.main}
\lim_{x\to\infty}\frac{\PP\left({\bf D}(T)\in xA\right)}{\PP({\bf X}\in xA)}=
\int_0^T
\E\left[\e^{-\alpha\xi(s)}\right]
\lambda(\dd s).
\eeam
This relation separates the asymptotic behavior of the multidimensional
aggregate claims into the set-based tail behavior of a single claim vector
and an explicit factor determined by the claim-arrival and investment
processes. More importantly, the set-based formulation allows us to
accommodate dependence among different claim vectors, rather than requiring
the successive claims to be independent. Moreover, the investment return
process $\xi$ need not be a L\'evy process, and the counting process $\{N(t),\, t\geq 0\}$ is not necessarily a renewal process. This more general treatment of the investment process is made possible by imposing a suitable moment condition of order strictly larger than the tail index of claim vectors, or, by a tail-negligibility condition on the stochastic discount factors. For the counting process $\{N(t),\, t\geq 0\}$, we assume that it satisfies a mild moment condition.

The rest of the paper is organized as follows.
Section~\ref{sec.CKP28T.2} introduces the required one-dimensional
heavy-tailed distribution classes and the set-based $\operatorname{MRV}$ and set-based $\operatorname{MRV}^\ast$, together with their
relationship to classical $\operatorname{MRV}$. We also provide some closure properties of our new classes, one of which shows that the class of strongly regularly varying distributions is not closed with respect to convolution.
Section~\ref{sec.CKP28T.3} presents the assumptions, main results, and  proofs.
Section~\ref{sec.CKP.4} provides a numerical study examining the accuracy
of the asymptotic approximation and its sensitivity to model parameters.
\section{Preliminaries} \label{sec.CKP28T.2}

Hereafter, unless otherwise stated, all asymptotic relations are understood as $\xto$. For two positive univariate functions $f$ and $g$, we write $f(x)\sim c\,g(x)$ for some $c\in(0,\infty)$, $f(x)=o[g(x)]$,
$f(x)=O[g(x)]$, or $f(x)\asymp g(x)$ if it holds respectively,
\beao
\lim \frac{f(x)}{g(x)}=c,\
\lim \frac{f(x)}{g(x)}=0,\
\limsup \frac{f(x)}{g(x)}<\infty,\
0<\liminf \frac{f(x)}{g(x)}
\leq \limsup \frac{f(x)}{g(x)}<\infty.
\eeao

For a set $B$, we denote by $\overline{B}$ its closure, by $\partial B$ its boundary, and by $B^c$ its complement.
%\textcolor{red}{If ${\bf f}^*$, ${\bf g}^*$ are two positive $d$-variate functions and $B \in \bbr^d:=(-\infty,\infty)^d$ with ${\bf 0} \notin \overline{B}$, with ${\bf 0} =(0,\,\ldots,\,0)^{\top}$, then the previous asymptotic notations remain in tact, for example we write $f^*(x\,B) \sim c\,g^*(x\,B)$ with  $c \in (0,\,\infty)$, if}
%\beao
%\lim \dfrac{f^*(x\,B)}{g^*(x\,B)} =c\,.
%\eeao
Further, for two real numbers $a$ and $b$, we write $a\vee b:=\max\{a,b\}$. Throughout, vectors are $d$-dimensional, with $d\in\bbn$, and are written in boldface; for example, ${\bf x}=(x_1,\ldots,x_d)^{\top}$. The vector operations are defined componentwise, namely
${\bf x}\pm{\bf y}=(x_1\pm y_1,\ldots,x_d\pm y_d)^{\top}$,
while scalar multiplication by $b>0$ is given by
$b{\bf x}=(bx_1,\ldots,bx_d)^{\top}$.

If a random variable (or, random vector) $Z$ has distribution $V$, we write $Z\stackrel{d}{\sim}V$. For two distributions $V_1$ and $V_2$, we denote their convolution by $V_1*V_2$. If $V$ is a one-dimensional distribution, its tail is denoted by $\bV(x):=1-V(x)$, $x\in\bbr$.

\subsection{One-dimensional heavy-tailed distributions} \label{subsec.CKP28T.2.1}

We now introduce some preliminary concepts concerning one-dimensional heavy-tailed distributions that will be used later. 
%Throughout this subsection, {\color{blue}all distributions are assumed to be defined on  $x\in\bbr^{+}$}.

We say that a distribution $V$ defined on $\mathbb R$ is heavy-tailed, denoted by $V\in\mathcal{K}$, if
\beao
\widehat{V}(\vep):=\int_{-\infty}^{\infty} \e^{\vep x}\,V(\dd x)=\infty
\eeao
for every $\vep>0$. We denote by $\mathcal{K}^c$ the class of light-tailed distributions.

We say that a distribution $V$ defined on $\mathbb R$ belongs to the class of exponential-like tailed distributions with parameter $\alpha\geq0$, denoted by $V\in\mathcal{L}(\alpha)$, if it holds for every $y>0$ that
\[
\lim \dfrac{\bV(x-y)}{\bV(x)}=\e^{\alpha y}.
\]
When $\alpha=0$, the class $\mathcal{L}(0)\equiv\mathcal{L}$ is the class of long-tailed distributions, and
\(
\mathcal{L}\subsetneq\mathcal{K}.
\)
When $\alpha>0$, we have
\(
\mathcal{L}(\alpha)\subsetneq\mathcal{K}^c.
\)

We say that a distribution $V$ defined on $\mathbb R_{+}$ belongs to the class of convolution-equivalent distributions with parameter $\alpha\geq0$, denoted by $V\in\mathcal{S}(\alpha)$, if the following conditions hold:
\begin{enumerate}
\item[(i)] $V\in\mathcal{L}(\alpha)$;
\item[(ii)]
\[
\lim \dfrac{\overline{V*V}(x)}{\bV(x)}=2\,\widehat{V}(\alpha)<\infty.
\]
\end{enumerate}
 For a distribution $V$ defined on $\mathbb R$, we still say that it belongs to the class of convolution-equivalent distributions with parameter $\alpha\geq0$ if $V_+\in\mathcal S(\alpha)$, where $V_+(x)=V(x)\boldsymbol{1}_{\{x\geq 0\}}$, see \cite[Cor. 2.1(i)]{pakes:2004}.

When $\alpha=0$, the class $\mathcal{S}(0)\equiv\mathcal{S}$ is the class of subexponential distributions, and $\mathcal{S}\subsetneq\mathcal{L}$. When $\alpha>0$, we have $\mathcal{S}(\alpha)\subsetneq\mathcal{L}(\alpha)\subsetneq\mathcal{K}^c$.

The classes $\mathcal{S}$, $\mathcal{L}$, and $\mathcal{K}$ were introduced by
\cite{chistyakov:1964}, while, for $\alpha>0$, the classes
$\mathcal{S}(\alpha)$ and $\mathcal{L}(\alpha)$ were subsequently introduced by
\cite{chover:ney:wainger:1973a} and
\cite{chover:ney:wainger:1973b}. These early developments were motivated
by applications to branching processes. For further properties of these
distribution classes, we refer the reader to
\cite{embrechts:klueppelberg:mikosch:1997,borovkov:borovkov:2008,
foss:korshunov:zachary:2013,leipus:siaulys:konstantinides:2023}.

Further, the class of consistently varying distributions $\mathcal{C}$, represents an important subclass of $\mathcal{S}$. We say that a distribution $V$ defined on $\mathbb R$ belongs to the class 
$\mathcal{C}$ if it holds that
\[
\lim_{b\uparrow 1}\limsup
\dfrac{\bV(bx)}{\bV(x)}=1\,.
\]
The class $\mathcal{C}$ contains the class of regularly varying distributions. We say that a distribution $V$ is regularly varying with tail index $\alpha\in[0,\infty)$, denoted by $V\in\mathcal{R}_{-\alpha}$, if
\[
\lim\dfrac{\bV(bx)}{\bV(x)}
=b^{-\alpha}
\]
for every $b>0$. The case $\alpha=0$ corresponds to the class of distributions with slowly varying tails. The classes $\mathcal{R}_{-\alpha}$ and $\mathcal{L}(\alpha)$ are related through the following transformation,
\beam \label{eq.CKP.2.5}
\bV(x)=\frac{\bU(\e^x)}{\bU(0)}\,,
\eeam
where $U$ is a proper distribution on $\bbr$. If $Z\stackrel{d}{\sim}U$, then $V$ is the conditional distribution of $\log Z$ given $Z>0$. It is easy to see that, for every $\alpha\geq 0$, $U\in\mathcal{R}_{-\alpha}$ if and only if $V\in\mathcal{L}(\alpha)$.

Motivated by this observation, \cite[Def. 2.1]{li:tang:2015} introduced the class of strongly regularly varying distributions. 

We say that a distribution $U$ is strongly regularly varying with tail index $\alpha\in[0,\infty)$, denoted by $U\in\mathcal{R}_{-\alpha}^*$, if the distribution $V$ defined through \eqref{eq.CKP.2.5} belongs to $\mathcal{S}(\alpha)$.
Since $\mathcal{S}(\alpha)\subsetneq\mathcal{L}(\alpha)$, it follows from \eqref{eq.CKP.2.5} that
$\mathcal{R}_{-\alpha}^*\subsetneq\mathcal{R}_{-\alpha}$ for every $\alpha\geq 0$. Moreover, if $Z\stackrel{d}{\sim}U\in\mathcal{R}_{-\alpha}^*$, then
\[
\E[Z_{+}^\alpha]<\infty\,,
\]
because
\[
\E[Z_{+}^\alpha]=\bU(0)\,\widehat{V}(\alpha)<\infty.
\]

%The following result follows directly from \eqref{eq.CKP.2.5} and the corresponding properties of the classes $\mathcal{S}(\alpha)$ and $\mathcal{L}(\alpha)$.
The following proposition follows directly from \eqref{eq.CKP.2.5}, together with Lemma 2.4(i) in \cite{pakes:2004} for part~(i)  and Lemma 2.6 in \cite{watanabe:2008} for part~(ii).

\bpr \label{pr.CKP.2.1}
Let $U_1$ and $U_2$ be two distributions on $\bbr$.
\begin{enumerate}
\item[(i)]
If $U_1\in\mathcal{R}_{-\alpha}^*$ and
$\bU_2(x)\sim c\,\bU_1(x)$ for some $c\in(0,\infty)$, then
$U_2\in\mathcal{R}_{-\alpha}^*$.
\item[(ii)]
If $U_1\in\mathcal{R}_{-\alpha}^*$,
$U_2\in\mathcal{R}_{-\alpha}$, and
$\bU_1(x)\asymp\bU_2(x)$, then
$U_2\in\mathcal{R}_{-\alpha}^*$.
\end{enumerate}
\epr

As noted by \cite{li:tang:2015}, in view of the transformation \eqref{eq.CKP.2.5}, sufficient conditions for a distribution to belong to the class $\mathcal{R}^*_{-\alpha}$ can be obtained from corresponding sufficient conditions for membership in $\mathcal{S}(\alpha)$. For such conditions, we refer to \cite{cline:1986,kluppelberg:1989,watanabe:2019}. For further discussion of the class $\mathcal{R}^*_{-\alpha}$, see \cite{li:tang:2015,cui:omey:wang:wang:2018}.

Concluding, the following inclusions are well-known, see for example \cite[Ch. 2]{leipus:siaulys:konstantinides:2023}
\[
\mathcal{R}^* \subsetneq \mathcal{R} \subsetneq \mathcal{C} \subsetneq \mathcal{S} \subsetneq \mathcal{L} \subsetneq \mathcal{K} \,,
\]
where $\mathcal{R}^* := \bigcup_{\alpha \geq 0} \mathcal{R}^*_{-\alpha}$, $\mathcal{R} := \bigcup_{\alpha\geq 0} \mathcal{R}_{-\alpha}$.

Finally, we close this subsection with the Matuszewska indices, which play an important role in the characterization of several subclasses of $\mathcal{K}$. The lower and upper Matuszewska index of a distribution $V$ are defined as follows:
\beao
J_V^- = -\lim_{\yto} \dfrac{\log \bV^*(y)}{\log y}\,, \qquad J_V^+ = -\lim_{\yto} \dfrac{\log \bV_*(y)}{\log y}\,,
\eeao
where 
\beao
\bV_*(y) = \liminf \dfrac{\bV(y\,x)}{\bV(x)}\,, \qquad \bV^*(y) = \limsup \dfrac{\bV(y\,x)}{\bV(x)}\,.
\eeao
We note that if $V \in \mathcal{R}_{-\alpha}$ with  $\alpha \in [0,\,\infty)$, then $J_V^- =J_V^+ =\alpha$. By \cite[Lem. 3.5]{tang:tsitsiashvili:2003}, for a distribution $V$, if $J_V^+<\infty$, then for every $q>J_V^+$ it holds that
\beam \label{eq.CKP.2.12}
x^{-q}=o[\bV(x)]\,.
\eeam
See  \cite[Sub. 2.1.2]{bingham:goldie:teugels:1987} for more discussions about Matuszewska indices.

\subsection{Set-based $\operatorname{MRV}$ and set-based $\operatorname{MRV}^\ast$} \label{subsec.CKP28T.2.2} 

In this subsection, we introduce a set-based notion of $\operatorname{MRV}$. We also compare this notion with the classical $\operatorname{MRV}$ based on vague convergence of measures.

In order to define set-based $\operatorname{MRV}$ and set-based $\operatorname{MRV}^\ast$, we need to introduce the following family of sets introduced in \cite{samorodnitsky:sun:2016}: 
\[
\mathscr{R}=\left\{
A\subsetneq\bbr^d:
A\ {\text{is open and increasing}},\
A^c\ {\text{is convex}},\
{\bf 0}\notin\overline{A}
\right\},
\]
where a set $A$ is called increasing if ${\bf x}\in A$ and ${\bf y}\in\bbr_+^d$ imply ${\bf x}+{\bf y}\in A$. We note that $\mathscr{R}$ is closed under positive scalar multiplication; namely, for every $A\in\mathscr{R}$ and $b>0$, we have $bA\in\mathscr{R}$. Sets $A_1,\, A_2$ of \eqref{eq.CKP.1.4b} and \eqref{eq.CKP.1.4a}, are two important sets belonging to family $\mathscr{R}$, see \cite{samorodnitsky:sun:2016} for more examples on such sets.

Let ${\bf Y}\stackrel{d}{\sim}G$ be defined on $\bbr_+^d$. For any $A\in\mathscr{R}$, define
\beam \label{eq.CKP.2.9}
Y_A:=\sup\{u>0:{\bf Y}\in uA\}\,.
\eeam
Then $Y_A$ has a proper distribution $G_A$, whose tail is given by
\beam \label{eq.CKP.2.10}
\bG_A(x)=
\PP({\bf Y}\in xA)=
\PP\left(\sup_{{\bf p}\in I_A}{\bf p}^{\top}{\bf Y}>x\right),
\eeam
for every $x>0$. The existence of the index set $I_A$ for every $A\in\mathscr{R}$ follows from  \cite[Lem. 4.3(c)]{samorodnitsky:sun:2016}, while the fact that $G_A$ is a proper distribution follows from \cite[Lem. 4.5]{samorodnitsky:sun:2016}.
 
 By \cite{samorodnitsky:sun:2016}, for a fixed set $A\in\mathscr{R}$ we say that a distribution $G$ is multivariate subexponential on $A$, symbolically $G\in\mathcal{S}_A$, if $G_A\in\mathcal{S}$. Similarly, \cite{konstantinides:passalidis:2024g} introduced the class $\mathcal{C}_A$ of consistently varying distributions, as the class which contains the distributions $G$ such that $G_A\in\mathcal{C}$. For examples of classes $\mathcal{S}_A,\,\mathcal{C}_A$, see \cite[Sect. 4]{samorodnitsky:sun:2016}, \cite[Sect. 4]{konstantinides:liu:passalidis:2025}.

We now introduce the following concepts on set-based $\operatorname{MRV}$ and set-based $\operatorname{MRV}^\ast$.
\begin{definition}
Let $G$ be a $d$-dimensional distribution on $\bbr_+^d$, and let $A\in\mathscr{R}$.
\begin{enumerate}
\item[(i)]
We say that $G$ is $A$-based $\operatorname{MRV}$ with index $\alpha\geq 0$, denoted by
$G\in\mathcal{R}_{-\alpha,A}$, if
\(
G_A\in\mathcal{R}_{-\alpha}.
\)
\item[(ii)]
We say that $G$ is $A$-based $\operatorname{MRV}^\ast$ with index $\alpha\geq 0$, denoted by
$G\in\mathcal{R}_{-\alpha,A}^*$, if
\(
G_A\in\mathcal{R}_{-\alpha}^*.
\)
\end{enumerate}
\end{definition}

Directly from one dimensional ordering of heavy-tailed distributions, for any fixed set $A\in\mathscr{R}$ it holds that
\beam\label{multi-incl}
\mathcal{R}_A^* \subsetneq \mathcal{R}_A \subsetneq \mathcal{C}_A \subsetneq \mathcal{S}_A\,,
\eeam
where $\mathcal{R}_A^* := \bigcup_{\alpha \geq 0}\, \mathcal{R}_{-\alpha,A}^*$, $\mathcal{R}_A := \bigcup_{\alpha\geq 0} \mathcal{R}_{-\alpha,A}$.

In contrast to regular variation on a fixed set $A\in\mathscr{R}$, the classical notion of $\operatorname{MRV}$ was introduced in \cite{haan:resnick:1981}. Let
\(
\mathbb{E}:=[0,\infty]^d\setminus\{{\bf 0}\}.
\)
We say that a $d$-dimensional distribution $G$ is $\operatorname{MRV}$ with index $\alpha>0$, if there exist a distribution $V\in\mathcal{R}_{-\alpha}$ and a Radon measure $\mu$, that is a non-degenerate at zero on $\mathbb{E}$, such that
\beam \label{eq.CKP.2.11}
\frac{1}{\bV(x)}
\PP\left({\bf Y}/x\in \cdot\right)
\xrightarrow{v}
\mu(\cdot),
\eeam
where $\xrightarrow{v}$ denotes the vague convergence, and in this case we write $G\in\operatorname{MRV}(\alpha,V,\mu)$. The vague convergence, implies that \eqref{eq.CKP.2.11}
holds (as $x\to\infty$) for every Borel set $B\subseteq\mathbb{E}$ that is bounded away from the origin and satisfies $\mu(\partial B)=0$.
The limit measure $\mu$ is homogeneous of order $-\alpha$, that is,
\beao
\mu(cB)=c^{-\alpha}\mu(B),\qquad c>0,
\eeao
for every Borel set $B\subseteq\mathbb{E}$.

\bre \label{rem.CKP.MRVstar}
A stronger version of $\operatorname{MRV}$ is obtained by strengthening the tail condition on the reference distribution $V$. More precisely, if relation \eqref{eq.CKP.2.11} holds with
$V\in\mathcal{R}_{-\alpha}^*$, we say that $G$ is $\operatorname{MRV}^\ast$ with index $\alpha>0$, and write
\[
G\in\operatorname{MRV}^*(\alpha,V,\mu).
\]
Since $\mathcal{R}_{-\alpha}^*\subsetneq\mathcal{R}_{-\alpha}$, every $\operatorname{MRV}^\ast$ distribution is $\operatorname{MRV}$.
\ere

\bre \label{rem.CKP.mu.positivity}
If ${\bf Y}\stackrel{d}{\sim}G\in\operatorname{MRV}(\alpha,V,\mu)$, by the fact that $\mu$ is a Radon measure non-degenerate at zero on $\mathbb{E}$, we have that at least one of the components of ${\bf Y}$ has regularly varying distribution with parameter $\alpha>0$, and it is strong tail equivalent with $V$. By this fact, we can see easily that $\mu(A_i)>0$ for $i=1,2$. However $\mu(A)>0$ is not true for any $A\in\mathscr{R}$  (see, Example \ref{eg.CKP.eg0} below), as claimed by \cite[Prop. 4.14]{samorodnitsky:sun:2016} and \cite[Prop. 2.1]{konstantinides:passalidis:2024g}, and hence their inclusions about the order of multivariate heavy-tailed distribution classes need the extra condition of positive $\mu$. We also note that for every $A\in\mathscr{R}$ it holds $\mu(A)<\infty$, because ${\bf 0}\notin \overline{A}$, and $\mu$ is homogeneous Radon measure.
\ere

We next compare the classical  $\operatorname{MRV}$ with $\mathcal{R}_{-\alpha,A}$, and  $\operatorname{MRV}^\ast$ with $\mathcal{R}_{-\alpha,A}^*$, on a fixed set $A\in\mathscr{R}$. The proof is a direct application of the definition of the classical  $\operatorname{MRV}$ and is therefore omitted.

\bpr \label{pr.CKP.MRVcompare}
Let $A\in\mathscr{R}$.
\begin{enumerate}
\item[(i)]
If $G\in\operatorname{MRV}(\alpha,V,\mu)$ 
with $\alpha>0$ and $\mu(A)>0$, then
\(
G\in\mathcal{R}_{-\alpha,A}.
\)
\item[(ii)]
If $G\in\operatorname{MRV}^*(\alpha,V,\mu)$ 
with $\alpha>0$ and $\mu(A)>0$, then
\(
G\in\mathcal{R}_{-\alpha,A}^*.
\)
\end{enumerate}
\epr
%From the last proposition and \eqref{multi-incl}, we have that for any $A\in\mathscr{R}$ such that $\mu(A)>0$ it holds:
%{\color{blue}\begin{align}
%    \label{MRV-incl}
%&\operatorname{MRV}^\ast \subsetneq \operatorname{MRV},\quad \operatorname{MRV}\subsetneq\mathcal{R}_A,\quad\operatorname{MRV}^{\ast}\subsetneq\mathcal{R}^{\ast}_A, \quad \mathcal{R}_A^* \subsetneq \mathcal{R}_A \subsetneq \mathcal{C}_A \subsetneq \mathcal{S}_A\,,
%\end{align}}
%where $\operatorname{MRV}^\ast$ and $\operatorname{MRV}$ are the unions of all multiavariate (stronly) regular varying distributions with index $\alpha>0$ and some Radon measure $\mu$.

The condition $\mu(A)>0$ in Proposition~\ref{pr.CKP.MRVcompare} is essential.
Without this condition, the classical  $\operatorname{MRV}$ and
$A$-based $\operatorname{MRV}$ are in general incomparable; the same is true
for the classical  $\operatorname{MRV}^\ast$ and $A$-based $\operatorname{MRV}^\ast$.
The following two examples illustrate the two directions. 

Let
\[
A_3=\left\{(x_1,x_2)\in\bbr^2:x_2>1\right\}\in\mathscr R.
\]
The following example shows that a classical $\operatorname{MRV}$ distribution may not have a heavy-tailed distribution $F_{A_3}$ when $\mu(A_3)=0$.
\begin{example}\label{eg.CKP.eg0}
Let $Y$ and $S$ be two positive random variables such that
\[
\overline V(x):=\PP(Y>x)=x^{-\alpha},\qquad x\geq1,
\]
and
\[
\bG(x):=\PP(S>x)=\mathrm{e}^{-x},
\qquad x\geq0.
\]
Then $\bG(x)=o(x^{-\alpha})$, and
$G$ is not heavy-tailed. Define
\[
{\bf X}=
\begin{cases}
(Y,0), & \text{with probability }1/2,\\
(0,S), & \text{with probability }1/2.
\end{cases}
\]
Let ${\bf X}\stackrel{d}{\sim}F$. For
${\bf t}=(t_1,t_2)$ with $t_1,t_2>0$, ${\bf 1}=(1,1)$:
\[
1-F(x{\bf t})=\frac12(xt_1)^{-\alpha}+\frac12\bG(xt_2)\sim\frac12(xt_1)^{-\alpha},
\]
whereas
\[
1-F(x{\bf 1})=\frac12x^{-\alpha}+\frac12\bG(x)\sim \frac12x^{-\alpha}.
\]
Hence,
\[
\lim\frac{1-F(x{\bf t})}{1-F(x{\bf 1})}= t_1^{-\alpha}.
\]
Thus, according to \cite[Th. 6.1]{resnick:2007}, $F\in\operatorname{MRV}(\alpha,V,\mu)$, where $\mu([{\bf 0},{\bf t}]^c)=t_1^{-\alpha}/2$, ${\bf t}=(t_1,t_2)\in[0,\infty)^{2}\setminus \{\bf 0\}$.
%Now take
%\[
%A_3=\left\{(x_1,x_2)\in\bbr^2:x_2>1\right\}\in\mathscr R.
%\]
Since $\mu$ is concentrated on the first coordinate axis, we have $\mu(A_3)=0$. On the other hand,
\[
\overline F_{A_3}(x)=\PP(X_2>x)=\frac12\overline G(x)=\frac12\mathrm e^{-x},\qquad x\ge0.
\]
Thus, $F_{A_3}$ is light-tailed and hence belongs to none of the heavy-tailed classes
$\mathcal R$, $\mathcal C$, $\mathcal S$, $\mathcal L$, and $\mathcal K$.
In particular,
\(
F\notin\mathcal R_{-\alpha,A_3}.
\)
\end{example}

Let 
\[
A^\prime_1=\left\{(x_1,x_2)\in\bbr^2:x_1\vee x_2>1\right\}\in\mathscr{R}.
\]
The next example shows that $F$ is $A^\prime_1$-based $\operatorname{MRV}$ but is not $\operatorname{MRV}$.
\begin{example}\label{eg.CKP.eg1}
Let $Y$ be a positive random variable with a continuous distribution
$H\in\mathcal{R}_{-\alpha}$, where $\alpha>0$, and assume that $Y\geq1$ almost surely. Fix $c>1$ and define the two-dimensional random vector ${\bf X}=(X_1,X_2)^{\top}$ by
\[
{\bf X}=
\begin{cases}
(Y,0)^{\top}, & c^{2k}\leq Y<c^{2k+1},\\
(0,Y)^{\top}, & c^{2k+1}\leq Y<c^{2k+2},
\end{cases}
\qquad k=0,1,2,\ldots.
\]
Since $X_1\vee X_2=Y$, we have
\[
\bF_{A^\prime_1}(x)=
\PP({\bf X}\in xA^\prime_1)=
\PP(Y>x)=\bH(x).
\]
Hence,
\(
F\in\mathcal{R}_{-\alpha,A^\prime_1}.
\)
We now show that $F$ is not $\operatorname{MRV}$. Let ${\bf t}=(1,c)^{\top}$. Since
\[
1-F(x{\bf 1})=
\PP(X_1>x\ \text{or}\ X_2>x)=
\PP(Y>x)=\bH(x),
\]
for $x=c^{2n}$ we obtain
\[
1-F(c^{2n}{\bf t})=\PP(X_1>c^{2n}\ \text{or}\ X_2>c^{2n+1})=\bH(c^{2n}),
\]
and therefore
\[
\frac{1-F(c^{2n}{\bf t})}
     {1-F(c^{2n}{\bf 1})}=1.
\]
On the other hand, for $x=c^{2n+1}$,
\[
1-F(c^{2n+1}{\bf t})=\PP(X_1>c^{2n+1}\ \text{or}\ X_2>c^{2n+2})=
\bH(c^{2n+2}),
\]
and hence, since $H\in\mathcal{R}_{-\alpha}$,
\[
\frac{1-F(c^{2n+1}{\bf t})}
     {1-F(c^{2n+1}{\bf 1})}=
\frac{\bH(c^{2n+2})}{\bH(c^{2n+1})}
\longrightarrow
c^{-\alpha}.
\]
Since $c^{-\alpha}\neq1$, the limit
\[
\lim
\frac{1-F(x{\bf t})}{1-F(x{\bf 1})}
\]
does not exist. According to \cite[Th. 6.1]{resnick:2007}, $F$ is not $\operatorname{MRV}$.
\end{example}
\begin{remark}
The constructions in Examples~\ref{eg.CKP.eg0} and \ref{eg.CKP.eg1}
can be adapted to the strongly regularly varying setting by choosing
the regularly varying baseline distributions from the corresponding
strongly regularly varying class. More precisely, in
Example~\ref{eg.CKP.eg0}, the distribution $V$ may be chosen from
$\mathcal R^*_{-\alpha}$ while keeping $G$ as an exponential distribution (or more general, $G$ can be choosen as any distribution from $\mathcal{K}^c$), whereas in
Example~\ref{eg.CKP.eg1}, one may simply take
$H\in\mathcal R^*_{-\alpha}$. Consequently, without the condition $\mu(A)>0$, the classical  $\operatorname{MRV}^\ast$ and
$A$-based $\operatorname{MRV}^\ast$ are, in general, incomparable.
\end{remark}

\begin{remark}
Now we can revised the inclusions of \cite{samorodnitsky:sun:2016, konstantinides:passalidis:2024g}. From Proposition \ref{pr.CKP.MRVcompare}, Example \ref{eg.CKP.eg0} and \eqref{multi-incl}, we have that for any $A\in\mathscr{R}$ such that $\mu(A)>0$, it holds that
\beam\label{MRV-incl}
\operatorname{MRV}\subsetneq \mathcal{R}_A \subsetneq \mathcal{C}_A \subsetneq \mathcal{S}_A\,,
\eeam
and
\beam\label{MRV*-incl}
\operatorname{MRV*}\subsetneq \mathcal{R}_A^* \subsetneq \mathcal{R}_A \subsetneq \mathcal{C}_A \subsetneq \mathcal{S}_A\,,
\eeam
where $\operatorname{MRV}^*$ and $\operatorname{MRV}$ are the unions of all multiavariate (stronly) regular varying distributions with index $\alpha>0$ and some Radon measure $\mu$.
\end{remark}

\subsection{Some closure properties of the new classes} \label{subsec.CKP28T.2.3}

In this subsection we provide some properties for the new classes $\mathcal{R}_{-\alpha,A}^*,\, \mathcal{R}_{-\alpha,A}$ and $\operatorname{MRV}^{*}$. More precisely, we want to find sufficient conditions for the convolution closure property of these classes, and we discuss the properties of linear combinations of their components. Also, we prove that classes $\mathcal{R}_{-\alpha,A}^*$ and $\operatorname{MRV}^{*}$ are not closed under convolution. 

First we note that Proposition \ref{pr.CKP.2.1} holds also for multivariate setup, for class $\mathcal{R}_{-\alpha,A}^*$, immediately by \eqref{eq.CKP.2.9}, \eqref{eq.CKP.2.10}. Similarly, if $F\in\mathcal{R}_{-\alpha,A}$, and $G(x\,A)\sim c\,F(x\,A)$, for some $c\in(0,\infty)$, then by the closure of regular variation with respect to strong tail equivalence (see for example, \cite[Prop. 3.3(i)]{leipus:siaulys:konstantinides:2023}), it holds $G\in\mathcal{R}_{-\alpha,A}$.

In this subsection, we consider two non-negative random vectors ${\bf X}  \stackrel{d}{\sim} F$ and ${\bf Y} \stackrel{d}{\sim} G$, that are mutually independent. To show the inclusion $F*G \in \mathcal{B}_A$, where $F*G(x\,A) := \PP({\bf X}+ {\bf Y} \in x\,A)$ for some $A \in \mathscr{R}$, and $\mathcal{B}$ stands for some distribution class, it is enough to prove that the transformation $Z_A:= \sup\{u>0 \;:\;{\bf X}+ {\bf Y} \in u\,A\} \stackrel{d}{\sim} H_A$ is such that $H_A \in \mathcal{B}$. The following proposition provides some sufficient conditions for the closure property of classes $\mathcal{R}_{-\alpha,\,A}$, $\mathcal{R}_{-\alpha,\,A}^*$, with respect to convolution.

\bpr \label{prop.CKP.2.A}
Let $A \in \mathscr{R}$ be a fixed set.
\begin{enumerate}
	\item
	If $F,\,G \in \mathcal{R}_{-\alpha,\,A}$, with $\alpha \geq 0$, then
	\beam \label{eq.CKP.2.A.1}
	F*G(x\,A) \sim F(x\,A)+ G(x\,A) \,,
	\eeam
	and further it holds $F*G \in \mathcal{R}_{-\alpha,\,A}$.
	\item
	If $F \in \mathcal{B}_A$, where $\mathcal{B}_A\in\left\{\mathcal{R}_{-\alpha,\,A},\,\mathcal{R}_{-\alpha,\,A}^* \right\}$, and $G(x\,A) = o[F(x\,A)]$, then
	\beam \label{eq.CKP.2.A.2}
	F*G(x\,A) \sim F(x\,A) \,,
	\eeam
	and $F*G \in \mathcal{B}_A$.
	\item
	If $F \in \mathcal{R}_{-\alpha,\,A}^*$, $G \in \mathcal{R}_{-\alpha,\,A}$, $G(x\,A) =O[F(x\,A)]$, then $F*G \in \mathcal{R}_{-\alpha,\,A}^*$.
\end{enumerate}
\epr

\pr~
We define $Y_A \stackrel{d}{\sim} G_A$, as in relations \eqref{eq.CKP.2.9}, \eqref{eq.CKP.2.10}, and correspondingly $X_A :=\sup\{u>0\;:\;{\bf X} \in u\,A\}\stackrel{d}{\sim} F_A$.
\begin{enumerate}
	\item
	Relation \eqref{eq.CKP.2.A.1} is implied by \cite[Th. 3.3(iii)]{konstantinides:passalidis:2024g}, since $\mathcal{R}_{-\alpha,\,A} \subsetneq \mathcal{C}_A$.
Further, for any $b>0$ it holds that
	\beam \label{eq.CKP.2.A.3} \notag
	&&\limsup \dfrac{\bH_A(b\,x)}{\bH_A(x)}=\limsup \dfrac{\PP({\bf X} + {\bf Y} \in b\,x\,A)}{\PP({\bf X} + {\bf Y} \in x\,A)}\leq\limsup \dfrac{F(b\,x\,A) + G(b\,x\,A)}{F(x\,A) + G(x\,A)} \\[2mm]
	&&\leq \max\left\{\limsup \dfrac{\bF_A(b\,x) }{\bF_A(x)} \,,\;\limsup \dfrac{\bG_A(b\,x)}{\bG_A(x)}  \right\}= b^{-\alpha}\,,
	\eeam
	where at the second step we used relation \eqref{eq.CKP.2.A.1} on the numerator and Bonferroni inequality on the denominator. By similar way, using the inverse arguments of \eqref{eq.CKP.2.A.3} for the numerator and denominator, we obtain for any $b>0$:
	\beam \label{eq.CKP.2.A.4} \notag
	&&\liminf \dfrac{\bH_A(b\,x)}{\bH_A(x)}\geq \liminf \dfrac{F(b\,x\,A) + G(b\,x\,A)}{F(x\,A) + G(x\,A)} \\[2mm]
	&&\geq \min\left\{\liminf \dfrac{\bF_A(b\,x) }{\bF_A(x)} \,,\;\liminf \dfrac{\bG_A(b\,x)}{\bG_A(x)}  \right\}= b^{-\alpha}\,.
	\eeam
	From relations \eqref{eq.CKP.2.A.3}, \eqref{eq.CKP.2.A.4} we find $H_A \in \mathcal{R}_{-\alpha}$, hence $F*G \in \mathcal{R}_{-\alpha,\,A}$. 
	\item
	Relation \eqref{eq.CKP.2.A.2} is implied by \cite[Lem. 3.4]{konstantinides:passalidis:yuan:2026}. Hence
	\beao
	\bH_A(x) \sim \bF_A(x)\,.
	\eeao
	From this last relation and the closure property of $\mathcal{B}$ with respect to strong tail equivalence, see \cite[Prop. 3.3(i)]{leipus:siaulys:konstantinides:2023} when $\mathcal{B} = \mathcal{R}_{-\alpha}$, and Prop. \eqref{pr.CKP.2.1}(i) when $\mathcal{B} = \mathcal{R}_{-\alpha}^*$, we find that $H_A \in \mathcal{B}$, hence $F*G \in \mathcal{B}_A$.
	\item
	From part (i) we obtain relation \eqref{eq.CKP.2.A.1}, and the inclusion $H_A \in \mathcal{R}_{-\alpha}$. Furthermore, since ${\bf X},\,{\bf Y}$ are non-negative, we obtain
	\beam \label{eq.CKP.2.A.5} 
	\bH_A(x) \geq \bF_A(x)\,,
	\eeam
	for any $x> 0$.
 On the other hand, since $G(x\,A) = O[F(x\,A)]$, and hence it holds $\bG_A(x) =O[\bF_A(x)]$, we find that there exist $C_1>0$, $x_0 >0$, such that it holds
	\beao
	\sup_{x > x_0} \dfrac{\bG_A(x)}{\bF_A(x)} \leq C_1\,.
	\eeao
	Furthermore, recall that the $G_A$, $F_A$ have infinite right endpoint, for any $x \leq x_0$, we obtain
	\beao
	\sup_{x \leq x_0} \dfrac{\bG_A(x)}{\bF_A(x)} \leq \sup_{x\leq x_0} \dfrac 1{\bF_A(x_0)} =: C_2\,.
	\eeao
	From these last two relations, choosing $C := C_1 \vee C_2$, and through relation \eqref{eq.CKP.2.A.1}, we have
	\beam \label{eq.CKP.2.A.6} 
	\bH_A(x) \sim \bF_A(x) + \bG_A(x) \leq (1+C)\,\bF_A(x) \,.
	\eeam
	
	From relations \eqref{eq.CKP.2.A.5}, \eqref{eq.CKP.2.A.6} we find $\bH_A(x) \asymp \bF_A(x)$, and since $F_A \in \mathcal{R}_{-\alpha}^*$, $H_A \in \mathcal{R}_{-\alpha}$, by Prop. \eqref{pr.CKP.2.1}(ii) we have $H_A\in\mathcal{R}_{-\alpha}^*$. Hence $F*G \in \mathcal{R}_{-\alpha,\,A}^*$.
\end{enumerate}
~\halmos

\bre \label{rem.CKP.2.2} 
Let ${\bf X} \stackrel{d}{\sim} F \in \operatorname{MRV}^*(\alpha,\,V,\,\mu)$. Then, for any ${\bf w} = (w_1,\,\ldots,\,w_d)\in (0,\infty)^d$ 
it holds that
\beao
\sum_{j=1}^d w_j\,X_j \sim F' \in \mathcal{R}_{-\alpha}^*\,.
\eeao
To check this, it is enough to consider the set $A_2$ and relation \eqref{eq.CKP.1.4a}.
However, the converse implication does not hold in general for $\operatorname{MRV}$; see \cite{hult:lindskog:2006}. More precisely,
\beam  \label{eq.CKP.2.17}
\forall \;{\bf w} \;:\;\sum_{j=1}^d w_j\,X_j \stackrel{d}{\sim} F' \in \mathcal{R}_{-\alpha}\; \not\Rightarrow \;F \in \operatorname{MRV}(\alpha,\,V,\,\mu)\,.
\eeam
 The converse implication also fails when $\mathcal{R}_{-\alpha}$ and $\operatorname{MRV}$ are replaced by $\mathcal{R}_{-\alpha}^*$ and $\operatorname{MRV}^*$, respectively, as the following example shows.
 \ere 
 
In opposite to the counterexample provided by \cite{hult:lindskog:2006}, here we assume independent components on ${\bf X}$. Such an example shows that the dependence is not the only factor from which caused implications like \eqref{eq.CKP.2.17}.
 
\begin{example}\label{eg.CKP.linear-SRV}
Fix $\alpha>0$. 
Let $X_1$ and $X_2$ be independent random variables with
distributions $F_1$ and $F_2$, whose tails are
\[
\overline F_1(x)=
\begin{cases}
1,&x<1,\\
h(x),&x\ge1,
\end{cases}
\]
\[
\overline F_2(x)=
\begin{cases}
1,&x<1,\\
r(x)h(x),&x\ge1.
\end{cases}
\]
where, for $x\geq1$, define
\[
h(x):=x^{-\alpha}(1+\log x)^{-2},\quad
\theta(x):=\log(1+\log x),\quad
r(x):=1+\frac12\sin\theta(x).
\]
It can be verified easily that they are valid continuous survival functions.
Now, we shall show that $F_1$ and $F_2$ are on $\mathcal{R}_{-\alpha}^*$.

Firstly, we verify regular variation of $F_1$ and $F_2$. For every fixed $t>0$,
\[
\theta(tx)-\theta(x)=\log\left(1+\frac{\log t}{1+\log x}\right)\to 0.
\]
Using the fact that $r(x)\geq1/2$, and the inequality
$|\sin u-\sin v|\leq|u-v|$, we obtain
\[
\left|\frac{r(tx)}{r(x)}-1\right|\leq|\theta(tx)-\theta(x)|\to 0.
\]
Furthermore,
\[
\frac{h(tx)}{h(x)}=t^{-\alpha}\left(\frac{1+\log(tx)}{1+\log x}\right)^{-2}\to t^{-\alpha}.
\]
Hence, $F_1,F_2\in\mathcal R_{-\alpha}$. 

We now show that $F_1\in\mathcal{R}_{-\alpha}^*$. Let $H$ be the distribution of $Z=\log X_1$. Then
\[
\overline H(x)=\e^{-\alpha x}(1+x)^{-2},\qquad x\ge0.
\]
By the fact that $F_1\in\mathcal{R}_{-\alpha}$ and by \eqref{eq.CKP.2.5}, it holds $H\in\mathcal L(\alpha)$. Moreover, integration of the tail gives
\[
\widehat H(\alpha):=\mathbb E[\e^{\alpha Z}]
=1+\alpha\int_0^\infty \e^{\alpha t}\overline H(t)\,\dd t
=1+\alpha\int_0^\infty(1+t)^{-2}\,\dd t
=1+\alpha<\infty.
\]
To check the convolution condition, it holds that
\[
\overline{H*H}(x)
=2\int_0^{x/2}\overline H(x-y)\,H(\dd y)+\overline H(x/2)^2.
\]
For $0\le y\le x/2$,
\[
0\le
\frac{\overline H(x-y)}{\overline H(x)}
=\e^{\alpha y}\left(\frac{1+x}{1+x-y}\right)^2
\le 4\e^{\alpha y}.
\]
Since $\int_0^\infty \e^{\alpha y}H(\dd y)=1+\alpha<\infty$,
the dominated convergence theorem gives
\[
\int_0^{x/2}\frac{\overline H(x-y)}{\overline H(x)}\,H(\dd y)
\longrightarrow\widehat H(\alpha).
\]
Also,
\[
\frac{\overline H(x/2)^2}{\overline H(x)}
=\frac{(1+x)^2}{(1+x/2)^4}\longrightarrow0.
\]
Consequently,
\[
\frac{\overline{H*H}(x)}{\overline H(x)}
\longrightarrow2\widehat H(\alpha)=2(1+\alpha).
\]
Thus $H\in\mathcal S(\alpha)$ and by \eqref{eq.CKP.2.5} we have $F_1\in\mathcal{R}_{-\alpha}^{*}$.

We also note further that
\[
\frac12\overline F_1(x)\leq\overline F_2(x)\leq\frac32\overline F_1(x),
\quad x\geq1.
\]
Hence, by Proposition \ref{pr.CKP.2.1} (ii), it holds $F_2\in\mathcal{R}_{-\alpha}^*$.

For $w_1,w_2>0$, let $F_{w_1,w_2}$ denote the distribution of
$w_1X_1+w_2X_2$. Since $X_1$ and $X_2$ are independent, by Proposition \eqref{prop.CKP.2.A}(1) it holds that
\[
\overline F_{w_1,w_2}(x)
\sim w_1 ^\alpha\overline  F_1(x)+w_2^\alpha\overline F_2(x)\asymp \overline  F_1(x).
\]
An application of Proposition~2.1(ii) gives
$F_{w_1,w_2}\in\mathcal R_{-\alpha}^{*}$.

However, denote the joint distribution of $(X_1, X_2)$ by $F$, we shall show that $F$ is not $\operatorname{MRV}$. In particular, take $\boldsymbol{t}=(2,1)^\top$.
By independence, for large $x$, we have
\begin{align*}
\frac{1-F(x\boldsymbol{t})}{1-F(x\boldsymbol{1})}&=
\frac{2^{-\alpha}\left(\frac{1+\log x}{1+\log(2x)}\right)^2+\left[1+\frac12\sin\!\bigl(\log(1+\log x)\bigr)\right]\left[1-\frac{(2x)^{-\alpha}}{(1+\log(2x))^2}\right]}{1+\left[1+\frac12\sin\!\bigl(\log(1+\log x)\bigr)\right]\left[1-\frac{x^{-\alpha}}{(1+\log x)^2}\right]}\\
&=\frac{1+2^{-\alpha}+\frac12\sin\!\bigl(\log(1+\log x)\bigr)}{2+\frac12\sin\!\bigl(\log(1+\log x)\bigr)}
+o(1),
\end{align*}
which does not have a limit when $x\to\infty$. By Theorem 6.1 in \cite{resnick:2007}, $F\notin\operatorname{MRV}$ and hence $F\notin\operatorname{MRV}^{*}$.
\end{example} 

We also note that, for marginals distributions in the  class $\mathcal{R}_{-\alpha}^*$, we can not take necessarily that their joint distribution belongs to $\mathcal{R}_{-\alpha,A}^*$.
In order to see this, we can take the set $A_2$, and independent marginals from class  $\mathcal{R}_{-\alpha}^*$. Then, due to the non-closure property of $\mathcal{R}_{-\alpha}^*$, with respect to convolution (see, Prop. \ref{prop.CKP.2.C}) we can find marginal distributions such that $F \notin \mathcal{R}^*_{-\alpha,A_2}$. A similar analytical example for multivariate subexponential distributions can be found in \cite[Exam. 2.1]{passalidis:2025}.

Let us continue now with some sufficient conditions for the closure property of $\operatorname{MRV}^*$, with respect to convolution, inspired by the corresponding properties of $\operatorname{MRV}$, that can be found in \cite[Cor. 3.2.35]{mikosch:wintenberger:2024}. Let us note that ${\bf X}+ {\bf Y} \stackrel{d}{\sim} F*G$, independently of the set of interest, as it was done previously in set-based classes. Furthermore, the following result, as also \cite[Cor. 3.2.35]{mikosch:wintenberger:2024}, continues to hold for the weakly dependent random vectors ${\bf X},\, {\bf Y}$, however for sake of compactness of the text, we still assume that these vectors are mutually independent.

\bpr \label{prop.CKP.2.B}
Let $\boldsymbol X\stackrel{d}{\sim}F$ and
$\boldsymbol Y\stackrel{d}{\sim}G$ be independent nonnegative random vectors. Let $V_1,V_2,V$ denote the distributions of $\|\boldsymbol X\|$, $\|\boldsymbol Y\|$, and $\|\boldsymbol X+\boldsymbol Y\|$, respectively, where $\|\cdot\|$ is the $L_1$ norm.
Suppose that $F\in\operatorname{MRV}^*(\alpha,V_1,\mu)$,
with $\alpha>0$, and
\beam \label{eq.CKP.2.A.7}
\lim_{x\to\infty}
\frac{\overline V_2(x)}{\overline V_1(x)}
=c^*\in[0,\infty).
\eeam
If $c^*>0$, assume additionally that
$G\in\operatorname{MRV}^*(\alpha,V_2,\nu)$;
if $c^*=0$, set $\nu=0$.
Then $V\in\mathcal R^*_{-\alpha}$ and
\beam \label{eq.CKP.2.A.8}
\frac{\PP((\boldsymbol X+\boldsymbol Y)/x\in\cdot)}
     {\overline V(x)}
\xrightarrow{v}
h(\cdot):=
\frac{\mu(\cdot)+c^*\nu(\cdot)}{1+c^*}.
\eeam
Consequently,
$F*G\in\operatorname{MRV}^*(\alpha,V,h)$.
\epr

\pr~
Applying \cite[Lem. 3.11]{jessen:mikosch:2006}
when $c^*>0$ and
\cite[Lem. 3.12]{jessen:mikosch:2006}
when $c^*=0$, together with \eqref{eq.CKP.2.A.7}, gives
\[
\frac{\PP((\boldsymbol X+\boldsymbol Y)/x\in\cdot)}
     {\overline V_1(x)}
\xrightarrow{v}
\mu(\cdot)+c^*\nu(\cdot).
\]
Since $\|\boldsymbol X+\boldsymbol Y\|=\|\boldsymbol X\|+\|\boldsymbol Y\|$, we have
\[
\overline V(x)\sim(1+c^*)\overline V_1(x).
\]
Hence, $V\in\mathcal R^*_{-\alpha}$ by
Proposition~\ref{pr.CKP.2.1}(i).
Combining the preceding two relations gives \eqref{eq.CKP.2.A.8}, and therefore $F*G\in\operatorname{MRV}^*(\alpha,V,h)$.
~\halmos

$\operatorname{MRV}$ with the same index does not, by itself, ensure closure under convolution. Furthermore, the positive answer, that is provided in Proposition \ref{prop.CKP.2.A} (i) for the $A$-based class $\mathcal{R}_{-\alpha,\,A}$, even under the condition $\mu(A)\vee \nu(A) >0$, for $A \in \mathscr{R}$, is not sufficient for a positive answer to this question, since the Dynkin's $\pi-\lambda$ theorem in the family of sets $ \mathscr{R}$, does not imply the validity on the set $\mathbb{E}$, which is needed for the $\operatorname{MRV}$ case.

As we shall see in the next proposition, the classes $\operatorname{MRV}^*$, $\mathcal{R}_{-\alpha,\,A}^*$, are not the closed with respect to convolution. To show this, it is enough to prove that the one-dimensional $\mathcal{R}_{-\alpha}^*$ is not closed with respect to convolution, since if $F \in \operatorname{MRV}^*(\alpha,\,V,\,\mu)$, with $d=1$, then it holds that
\beao
\frac{1}{\bV(x)}\PP(X \in x\,\cdot)\stackrel{v}{\rightarrow} \mu(\cdot)\,,
\eeao
and $\mu(\cdot)$ is Radon measure non-degerate at zero on $\bbr_{+}\setminus\{0\}$. The following proposition can be extended, by following similar techniques, to the case when the distribution have support on $\bbr$, through relation \eqref{eq.CKP.2.5}. However, for sake of concise of the text, we restrict ourselves on the case $\bbr_{+}$.

\bpr \label{prop.CKP.2.C}
For every $\alpha\geq0$, there exist distributions $U_1,U_2$ supported on $\bbr_+$ such that $U_1,U_2\in\mathcal{R}_{-\alpha}^*$ but $U_1*U_2\notin\mathcal{R}_{-\alpha}^*$.
\epr

\pr~
By \cite{leslie:1989,kluppelberg:villasenor:1991}, 
there exist $V_1,V_2\in\mathcal S(\alpha)$ such that $V_1*V_2\notin\mathcal S(\alpha)$.
Consequently, by \cite[Th. 1.1]{leipus:siaulys:2020}, $R:=(V_1+V_2)/2\notin\mathcal S(\alpha)$.

Let $Z_1,Z_2$ be independent random variables with distributions $V_1,V_2$, respectively, and let $U_i$ be the distribution of $\mathrm{e}^{Z_i}$, $i=1,2$.
Then $U_1,U_2\in\mathcal R_{-\alpha}^*$
by \eqref{eq.CKP.2.5}. Moreover,
Proposition~\ref{prop.CKP.2.A}(i) yields
\beam \label{eq.CKP.2.a}
\bV(x):=\overline{U_1*U_2}(\mathrm{e}^x)\sim \bU_1(\mathrm{e}^x)+\bU_2(\mathrm{e}^x)= \bV_1(x)+\bV_2(x)=2\overline R(x).
\eeam
Since $\mathcal S(\alpha)$ is closed under the strong tail equivalence, $V\notin\mathcal{S}(\alpha)$, that gives
$U_1*U_2\notin\mathcal R_{-\alpha}^*$.
~\halmos

\section{Main results and their proofs} \label{sec.CKP28T.3}

In this section, we present the main results and their proofs.

\subsection{Some assumptions} \label{subsec.CKP28T.3.1}
 Recall that the claim vectors $\{{\bf X}^{(i)}\,,\; i \in \bbn\}$, are identically distributed with distribution $F$. Similar to \eqref{eq.CKP.2.9} and \eqref{eq.CKP.2.10},  we define
\[
X_A^{(i)}:=\sup\{u>0:{\bf X}^{(i)}\in uA\}
\stackrel{d}{\sim}
F_A\,.
\]
The first assumption concerns the dependence among the claim vectors. Similar dependence conditions have been used in other risk models; see, for example, \cite{chen:konstantinides:passalidis:2025,konstantinides:passalidis:xu:2026,konstantinides:passalidis:yuan:2029}, where slightly more restrictive dependence structures were considered.
\begin{assumption} \label{ass.CKP.3.1}
Let $A \in \mathscr{R}$ be a fixed set. We assume that the claim vector sequence $\{ {\bf X}^{(i)}\,,\; i \in \bbn \}$ satisfies the  tail quasi-asymptotic independence on $A$, symbolically $TQAI_A$, namely, for any $i,\,j \in \bbn$ with $i \not= j$, it holds that  
\[
\lim_{x_i \wedge x_j \to \infty} \dfrac{\PP(X_A^{(i)}>x_i\,,\; X_A^{(j)}>x_j)}{\PP(X_A^{(i)}>x_i) + \PP(X_A^{(j)}>x_j) }=0\,.
\]
\end{assumption}

%\bre \label{rem.CKP.3.1}
%The  $TQAI$ for one-dimensional random variables, with $A =(1,\,\infty)$, was introduced in \yearcite{cheng:2014}.
%and it extended the quasi-asymptotic independence ($QAI$), found in \yearcite{chen:yuen:2009}. In the uni-variate case $QAI$, it was given by \eqref{eq.CKP.3.2} with $x_i= x_j$ for $A =(1,\,\infty)$. 
%In \yearcite{chen:konstantinides:passalidis:2025} were provided examples for $QAI_A$, for some sets  $A \in \mathscr{R}$. It is worth to mention, that $TQAI_A$, is one of the larger weak dependencies, that can be used to keep in tact the multivariate linear big jump principle, see related discussions in \yearcite{konstantinides:passalidis:xu:2026}.  
%\ere

The  $TQAI$ for one-dimensional random variables, with $A =(1,\,\infty)$, was introduced in \cite{cheng:2014}. The following example shows that the $TQAI_A$ condition allows nontrivial dependence among multivariate claim vectors. 

\begin{example}\label{eg.CKP.eg2}
Recall that $A_{1}$ defined by \eqref{eq.CKP.1.4b},
and let
\[
{\bf X}^{(i)}={\bf Y}^{(i)}+{\bf W},
\qquad i\in\bbn,
\]
where $\{{\bf Y}^{(i)},\,i\in\bbn\}$ are i.i.d. nonnegative random vectors and ${\bf W}$ is a nonnegative and non-degenerate at zero random vector, such that
\[
\bigvee_{k=1}^{d} W_k\leq M
\]
almost surely for some $M<\infty$. 
%We suppose that $\{({\bf Y}^{(i)},{\bf W}),\,i\in\bbn\}$ are i.i.d. coppies of $({\bf Y},{\bf W})$. 
Further, we assume that
\[
Z_i:=\bigvee_{k=1}^{d}Y_k^{(i)},\, i\in\mathbb{N}
\]
has common distribution $H\in\mathcal{L}$. Since the following inequalities hold almost surely
\[
Z_i\leq X_{A_1}^{(i)}
=\bigvee_{k=1}^{d}\bigl(Y_k^{(i)}+W_k\bigr)\leq Z_i+M,
\]
the random variables $\{X_{A_{1}}^{(i)},\,i\in\bbn\}$ are 
%identically distributed but
dependent through the common random vector ${\bf W}$. 
%Denoting their
%common distribution by $F_{A_1}$, we have
%\[
%\bF_{A_1}(x)\sim\bH(x).
%\]
Moreover, since $H\in\mathcal{L}$, for $i\neq j$, we obtain
%\[ 
%\PP\left(X_{A_1}^{(i)}>x_i,\,X_{A_1}^{(j)}>x_j\right)\leq\bH(x_i-M)\bH(x_j-M).
%\]
%Since $H\in\mathcal{L}$,
\[
\dfrac{\PP(X_A^{(i)}>x_i\,,\; X_A^{(j)}>x_j)}{\PP(X_A^{(i)}>x_i) + \PP(X_A^{(j)}>x_j) }\leq \frac{\bH(x_i-M)\bH(x_j-M)}{\bH(x_i)+\bH(x_j)}\lesssim\frac{\bH(x_i)\bH(x_j)}{\bH(x_i)+\bH(x_j)}
\longrightarrow0
\]
as $x_i\wedge x_j\to\infty$. Hence, the claim vectors
$\{{\bf X}^{(i)},\,i\in\bbn\}$ satisfy the $TQAI_A$ condition.
\end{example}

\bre \label{rem.CKP.3.1}
If in Assumption \ref{ass.CKP.3.1}, we take ``$x_i=x_j$" in the convergence, then we reduce to another popular weak dependence structure that named quasi-asymptotic independence on $A$ ($QAI_A$). We refer the reader to \cite[Sect. 2.2]{chen:konstantinides:passalidis:2025} and \cite[Exam. 2.2]{konstantinides:passalidis:yuan:2029} for examples and discussions on this dependence structure.
\ere

We next impose conditions on the stochastic process $\{\xi(t),\, t\geq 0\}$. Before that,  we need to introduce the random variable $\Theta^{*}(T):=\sup_{0\leq s\leq T} \e^{-\xi(s)} \stackrel{d}{\sim} H_T$.

\begin{assumption} \label{ass.CKP.3.2a}
Let $A \in \mathscr{R}$ be a fixed set and $T \in \Lambda$. We assume that $F \in \mathcal{R}_{-\alpha,A}$, with $\alpha>0$, and there exists some $q>\alpha$ such that
\beam \label{eq.CKP.3.3a}
\E\left[(\Theta^*(T))^q\right]<\infty\,.
\eeam
\end{assumption}

In the case of $F \in \mathcal{R}^*_{-\alpha,A}$, we can relax Assumption \ref{ass.CKP.3.2a} to the following one.

\begin{assumption} \label{ass.CKP.3.2}
Let $A \in \mathscr{R}$ be a fixed set and $T \in \Lambda$. We assume that $F \in \mathcal{R}^*_{-\alpha,A}$, with $\alpha>0$, and
\beam \label{eq.CKP.3.3}
\bH_T (x) = o\left[ \bF_A (x) \right]\,.
\eeam
\end{assumption}

\bre \label{rem.CKP.3.2}
Assumptions~\ref{ass.CKP.3.2a}, \ref{ass.CKP.3.2} are finite-time conditions and depend on the choice of $T\in\Lambda$. Assumption \ref{ass.CKP.3.2} requires the tail of the maximal stochastic discount factor over $[0,T]$ to be negligible compared with the tail of $F_A$. Moreover, (by the fact that $E[X_A^{\alpha}]<\infty$) \eqref{eq.CKP.3.3} implies
\begin{align}\label{eq.CKP.moment_condition}
\E\left[(\Theta^*(T))^\alpha\right]<\infty.
\end{align}
Consequently, since  for all $0\leq s\leq T$, it holds almost surely
\[
\e^{-\alpha\xi(s)}
\leq
(\Theta^*(T))^\alpha,
\]
we obtain from \eqref{eq.CKP.moment_condition} that
\beam \label{eq.CKP.3.a}
\sup_{0\leq s\leq T}
\E\left[\e^{-\alpha\xi(s)}\right]
<\infty.
\eeam

However, there exist many cases where Assumptions \ref{ass.CKP.3.2a},\ref{ass.CKP.3.2} satisfied for any $T\in\Lambda$. For example, when $\{\xi(t),\,t\geq0\}$ is a L\'evy process, Assumption~\ref{ass.CKP.3.2a} can be verified through an exponential moment condition. More precisely, suppose that
$F_A\in\mathcal{R}_{-\alpha}$ and define
\[
\phi(q):=\log\E\left[\e^{-q\xi(1)}\right].
\]
If $\phi(q)<\infty$ for some $q>\alpha$, then, for every fixed $T\in\Lambda$,
\[
\E\left[(\Theta^*(T))^q\right]<\infty;
\]
see, for example, Lemma 2.1 of \cite{griffin:2013}. In \cite[Sect. 4]{cheng:konstantinides:wang:2022} there exist a variety of (non-L\'{e}vy) commonly used financial models (such as Vasicek interest rate, Heston, and Stochastic volatility models) that satisfy Assumption \ref{ass.CKP.3.2a} for any fixed $T\in\Lambda$.
\ere

As we can see from Remark \ref{rem.CKP.3.2}, there exists a multitude of commonly used financial models $\{ \xi(t)\,,\;t \geq 0 \}$, for which relation \eqref{eq.CKP.3.3a} is satisfied, and even for any fixed $T\in\Lambda$. The main practical reason of the differentiation between Assumptions \ref{ass.CKP.3.2a} and \ref{ass.CKP.3.2}, is the more risky investments of the insurer. Thus, the following example, focuses in the differentiation between these two assumptions, and holds for any fixed $T\in\Lambda$.

\bexam \label{exam.CKP.3.A*}  
Let $\{ \xi(t)\,,\;t \geq 0 \}$ be a jump-diffusion  process of the form
\beam \label{eq.CKP.3.A.1}
\xi(t)=[\mu\,t + \sigma\,W(t)] - \sum_{i=1}^{N_J(t)} J_i =: B(t) - J(t)\,,
\eeam
with $t \geq 0$, where $\mu \in \bbr$ is the long-return rate, $\sigma >0$ is the diffusion coefficient and $\{ W(t)\,,\;t \geq 0 \}$ is a Wiener process. The jumps $\{ J_i\,,\; i \in \bbn \}$ are i.i.d. real-valued random variables with common distribution $G \in \mathcal{S}(\alpha)$, with $\alpha >0$, that arrive according to a homogeneous Poisson process $\{ N_J(t)\,,\;t \geq 0 \}$, with intensity $\lambda>0$. We suppose that the $\{ W(t)\,,\;t \geq 0 \}$, $\{ N_J(t)\,,\;t \geq 0 \}$ and $\{ J_i\,,\; i \in \bbn \}$ are mutually independent. Then for any fixed $T\in\Lambda$ we obtain
\beam \label{eq.CKP.3.A.2}
\E[(\Theta^*(T))^{\alpha}]< \infty\,, \qquad   \E[(\Theta^*(T))^{q}]= \infty\,,
\eeam 
for any $q> \alpha$.

Indeed, for the jump-diffusion process \eqref{eq.CKP.3.A.1},
conditioning on $N_J(1)$ and using independence give that
\[
\phi(q)
=\log \E[\e^{-q\xi(1)}]
=-\mu q+\frac12\sigma^2q^2+\lambda\bigl(\widehat G(q)-1\bigr),
\quad q>0,
\]
where $\widehat G(q):=\E[\e^{qJ_1}]$.
Since $G\in\mathcal S(\alpha)$, we have
\[
\phi(\alpha)<\infty\quad \text{and}\quad \phi(q)=\infty\quad\text{for}\quad q>\alpha.
\]
Then, according to Lemma 2.1 of \cite{griffin:2013}, we have $\E[(\Theta^*(T))^{\alpha}]< \infty$ for any fixed $T\in\Lambda$. Moreover,  for any $q>\alpha$ and any fixed $T\in\Lambda$, we have, by independence, that
\begin{align*}
\E[(\Theta^*(T))^q]&\ge\E\left[\e^{-q\xi(T)}\mathbf 1_{\{N_J(T)=1\}}\right]\\
&=\lambda T\exp\left\{-\lambda T-q\mu T+\frac12q^2\sigma^2T\right\}\E[\e^{qJ_1}]=\infty.
\end{align*}
This provides the second desired relation of \eqref{eq.CKP.3.A.2}.
\eexam

\bre \label{rem.CKP.3.2a}
Both of Assumptions \ref{ass.CKP.3.2a} and \ref{ass.CKP.3.2} implied that:
\begin{align}\label{eq.CKP.Breiman}
	\PP\left(\boldsymbol{X}\,\Theta^*(T)\in\,x\, A\right)
	\sim
	\PP(\boldsymbol{X}\,\in\,x\, A)\E\left[(\Theta^*(T))^\alpha\right],
\end{align}
provided $\boldsymbol{X}$ and $\Theta^*(T)$ are independent.
For the case of Assumption \ref{ass.CKP.3.2a}, this implied by the classical Breiman's theorem, see \cite{breiman:1965}, \cite[Cor. 3.6(iii)]{cline:samorodnitsky:1994}.

For the case of Assumption \ref{ass.CKP.3.2}, this follows by the alternative Breiman's analogue which given in \cite[Lem. 3.5]{li:tang:2015}. More precisely their Lemma 3.5, holds if for every $p \in (0,\,1)$ it holds $p\,F_A + (1-p)\,H_T \in \mathcal{R}^*_{-\alpha}$
As referred by \cite[p. 1804]{li:tang:2015} this is true in our case. Indeed, this can be verified by the transformation \eqref{eq.CKP.2.5}, \cite[Prop. 3.11(iv)]{leipus:siaulys:konstantinides:2023} and the fact that for every $p \in (0,\,1)$ it holds $p\,\bF_A(x) + (1-p)\,\bH_T(x)\sim p\,\bF_A(x)$.
\ere 

The following assumption is related to the common distribution of the claims and the counting process $\{N(t)\,,\;t \geq 0\}$.

\begin{assumption} \label{ass.CKP.3.3}
Let $A \in \mathscr{R}$ be a fixed set and $T\in\Lambda$. We assume that $F \in\mathcal{B}$, where  $\mathcal{B}\in \{\mathcal{R}_{-\alpha,A},\mathcal{R}^*_{-\alpha,A}\}$ and there exists $p> \alpha$ such that 
\beam \label{eq.CKP.3.A}
\E\left[ \left( N(T) \right)^{p +1} \right] < \infty.
\eeam
\end{assumption}

\bre \label{rem.CKP.3.2,5}
An equivalent condition for \eqref{eq.CKP.3.A} is the following
\[
\sum_{i=1}^{\infty} i^{p}\,\PP(\tau_i \leq T)< \infty\,.
\]
Condition \eqref{eq.CKP.3.A} has been used in several non-standard risk models, see for example \cite{gao:yang:2014,yang:su:2023}. For many quasi-renewal and inhomogeneous renewal processes Assumption \ref{ass.CKP.3.3} are satisfied for all $T\in\Lambda$, see \cite[Exam. 3.1, 3.2]{konstantinides:passalidis:yuan:2029}. By this, Remark \ref{rem.CKP.3.2} and Example \ref{exam.CKP.3.A*}, there exist a variety of not necessarily L\'{e}vy-Renewal models for which Assumption \ref{ass.CKP.3.2a}--\ref{ass.CKP.3.3} holds independently of the choice of $T\in\Lambda$.
\ere

\bre \label{rem.CKP.3.4}
If $\{{\bf X}^{(i)},\,i\in\bbn\}$, 
$\{\xi(t),\,t\geq0\}$, and 
$\{N(t),\,t\geq0\}$ are mutually independent, then relations \eqref{eq.CKP.3.a} and
\eqref{eq.CKP.3.A} imply that
\[
\sum_{i=1}^{\infty}\E\left[\Theta_i^\alpha(T)\right]<\infty.
\]
Indeed,
\begin{align*}
\sum_{i=1}^{\infty}\E\left[\Theta_i^\alpha(T)\right]
&=\sum_{i=1}^{\infty}\int_{0}^T\E\left[\e^{-\alpha\xi(s)}\right]\PP(\tau_i\in \dd s)
\\
&\leq\sup_{0\leq s\leq T}\E\left[\e^{-\alpha\xi(s)}\right]\sum_{i=1}^{\infty}\PP(\tau_i\leq T)
\\
&=\sup_{0\leq s\leq T}\E\left[\e^{-\alpha\xi(s)}\right]\E[N(T)]<\infty.
\end{align*}
\ere

\subsection{Main results}\label{subsec.CKP28T.3.1a}
The next two theorems give our main results of this paper.

\bth \label{th.CKP.3.1a}
Consider the risk model \eqref{eq.CKP.1.1}. Suppose that Assumptions~\ref{ass.CKP.3.1},  \ref{ass.CKP.3.2a}, and \ref{ass.CKP.3.3} hold for some  $T\in\Lambda$, and that 
	$\{{\bf X}^{(i)},\,i\in\bbn\}$,
	$\{\xi(t),\,t\geq0\}$, and 
	$\{N(t),\,t\geq0\}$ are mutually independent. Then,
	relation \eqref{eq.CPK.main} holds.
\ethe

In the next theorem, we limit the class of $F$ to the set-based $\operatorname{MRV}^\ast$ and we relaxed the moment condition of Assumption \ref{ass.CKP.3.2a} to relation \eqref{eq.CKP.3.3}.

\bth \label{th.CKP.3.1}
Consider the risk model \eqref{eq.CKP.1.1}. Suppose that Assumptions~\ref{ass.CKP.3.1}, \ref{ass.CKP.3.2}, and \ref{ass.CKP.3.3} hold  for some  $T\in\Lambda$, and that 
$\{{\bf X}^{(i)},\,i\in\bbn\}$,
$\{\xi(t),\,t\geq0\}$, and 
$\{N(t),\,t\geq0\}$ are mutually independent. Then
relation \eqref{eq.CPK.main} holds.
\ethe
 
\bre \label{rem.CKP.3.3}
Relation \eqref{eq.CPK.main} shows that the multivariate linear single-big jump principle remains valid in the present framework, which allows general counting processes and stochastic investment return processes as also weak dependence between the claim vectors. Theorem \ref{th.CKP.3.1} compared with the classical Breiman's theorem on Theorem \ref{th.CKP.3.1a}, does not require a moment condition of order strictly larger than $\alpha$ on the stochastic discount factors. Instead, the combination of
$F\in\mathcal{R}_{-\alpha,A}^*$ and the tail condition
\eqref{eq.CKP.3.3} allows us to apply the analogue of Breiman's theorem established in \cite[Lem. 3.5]{li:tang:2015}. In particular, as observed in Remark~\ref{rem.CKP.3.2}, these assumptions already imply
\eqref{eq.CKP.moment_condition}. We note that our Theorems are also new in one-dimensional case with $A=(1,\,\infty)$.
\ere

Although relation \eqref{eq.CPK.main} is explicit enough, there are some computational difficulties for $\PP({\bf X} \in x\,A)$, as $\xto$. Under different conditions with respect to dependence and distribution class of the marginals of the components of ${\bf X}$, and according to the choice of set $A \in \mathscr{R}$, the previous probability takes more explicit expressions, see for example, \cite[Sect. 4]{konstantinides:liu:passalidis:2025}.

In the classical  $\operatorname{MRV}$ (or, $\operatorname{MRV}^{\ast}$) setting, the asymptotic formula takes a simpler form. The following results follow directly from Theorems \ref{th.CKP.3.1a}, \ref{th.CKP.3.1} and Proposition \ref{pr.CKP.MRVcompare}.

\bco \label{cor.CKP.3.1a}
Under the conditions of Theorem~\ref{th.CKP.3.1a}, suppose further that
$F\in\operatorname{MRV}(\alpha,V,\mu)$ and that  $\mu(A)>0$. Then
\beam \label{eq.CKP.3.7}
\PP({\bf D}(T)\in xA)\sim\mu(A)\,\bV(x)\int_0^T\E\left[\e^{-\alpha\xi(s)}\right]\,\lambda(\dd s).
\eeam
\eco

\bco \label{cor.CKP.3.1}
Under the conditions of Theorem~\ref{th.CKP.3.1}, suppose further that
$F\in\operatorname{MRV}^*(\alpha,V,\mu)$ and that  $\mu(A)>0$. Then relation \eqref{eq.CKP.3.7} holds.
\eco

\bre \label{rem.CKP.3.5}
When $\{\xi(t),\,t\geq0\}$ is a L\'evy process, by the definition of the Laplace exponent $\phi$ given in Remark~\ref{rem.CKP.3.2}, we have:
\[
\E\left[\e^{-\alpha\xi(s)}\right]=\e^{s\phi(\alpha)},
\]
see for example \cite[Prop. 3.14]{cont:tankov:2004}. Hence, \eqref{eq.CPK.main} and \eqref{eq.CKP.3.7} reduce, respectively, to
\[
\PP({\bf D}(T)\in xA)
\sim
\PP({\bf X}\in xA)
\int_0^T 
\e^{s\phi(\alpha)}\,\lambda(\dd s),
\]
and
\[
\PP({\bf D}(T)\in xA)
\sim
\mu(A)\,\bV(x)
\int_0^T \e^{s\phi(\alpha)}\,\lambda(\dd s).
\]
\ere

\subsection{Proof of main Theorems} \label{subsec.CKP28T.3.2} 
In this section, we give the proofs of our main results, after some necessary preliminary lemmas.
The first lemma establishes the multivariate linear single big jump principle
for finite randomly weighted sums under the consistently varying-tailed
condition. Under the stronger assumption
$F\in\mathcal{R}_{-\alpha,A}^*$, it also yields an explicit asymptotic
form. 
%In this section we present the proof of Theorem \ref{th.CKP.3.1} after some preliminary lemmas. The first lemma shows the presence of the multivariate linear single big jump principle for randomly weighted sums, which is inspired by Theorem 1.2 in \yearcite{cheng:2014}. 

\ble \label{lem.CKP.3.1}
Let $A\in\mathscr{R}$ be a fixed set, and let
$\{\Theta_i,\,i\in\bbn\}$ be a sequence of nonnegative and non-degenerate to zero random variables, independent of
$\{{\bf X}^{(i)},\,i\in\bbn\}$. Suppose that Assumption~\ref{ass.CKP.3.1} holds and that
\beam \label{eq.CKP.3.7a}
\PP(\Theta_i>x)=o\left[\bF_A(x)\right],\quad
i\in\bbn.
\eeam
If $F\in\mathcal{C}_A$, then, for every fixed $n\in\bbn$,
\beam \label{eq.CKP.3.10}
\PP\left(\sum_{i=1}^n{\bf X}^{(i)}\Theta_i\in xA\right)
\sim
\sum_{i=1}^n
\PP\left({\bf X}^{(i)}\Theta_i\in xA\right).
\eeam
If, furthermore, $F\in\mathcal{R}_{-\alpha,A}^*$ for some $\alpha\geq 0$, then
\beam \label{eq.CKP.3.11}
\PP\left(\sum_{i=1}^n{\bf X}^{(i)}\Theta_i\in xA\right)
\sim
\PP({\bf X}\in xA)
\sum_{i=1}^n\E\left[\Theta_i^\alpha\right].
\eeam
\ele

\pr~
We first establish the upper bound of \eqref{eq.CKP.3.10}. We have
\beam  \label{eq.CKP.3.12}
&&\PP\left(\sum_{i=1}^n {\bf X}^{(i)}\,\Theta_i \in x\,A\right) \leq \PP\left( \sum_{i=1}^n X_{A}^{(i)}\,\Theta_i > x \right)\\[2mm] \notag
&& \sim \sum_{i=1}^n \PP\left( X_{A}^{(i)}\,\Theta_i > x\right) = \sum_{i=1}^n \PP\left( {\bf X}^{(i)}\,\Theta_i \in x\,A\right)\,,
\eeam
where at the first step we applied \cite[Prop. 2.4 ]{konstantinides:passalidis:2024g}. At the second step we applied Corollary 1.1 in \cite{cheng:2014}. 
Indeed, for each
$i\in\bbn$ and every $c>0$, choose $\delta>0$ such that
$\PP(\Theta_i>\delta)>0$.  Then, we have
\[
\PP\left(X_A^{(i)}\Theta_i>x\right)
\geq
\PP\left(\Theta_i>\delta\right)\bF_A(x/\delta).
\]
Since $F_A\in\mathcal{C}$, then
\[
\bF_A(cx)=
O\left[\bF_A(x/\delta)\right].
\]
Therefore,
\[
\PP(\Theta_i>cx)=o\left[\PP\left(X_A^{(i)}\Theta_i>x\right)\right]=o\left[\PP\left({\bf X}^{(i)}\Theta_i\in xA\right)\right],
\]
which verifies Assumption C of \cite{cheng:2014}.
For the lower bound in \eqref{eq.CKP.3.10}, since all the summands are
nonnegative and $xA$ is increasing, we have
\begin{align*}
&\PP\left(\sum_{i=1}^n{\bf X}^{(i)}\Theta_i\in xA\right)
\geq
\PP\left(\bigcup_{i=1}^n
\left\{{\bf X}^{(i)}\Theta_i\in xA\right\}\right)
\\ 
&\geq\sum_{i=1}^n
\PP\left({\bf X}^{(i)}\Theta_i\in xA\right)-\sum_{1\leq i<j\leq n}
\PP\left(
{\bf X}^{(i)}\Theta_i\in xA,\,
{\bf X}^{(j)}\Theta_j\in xA
\right).
\end{align*}
For any $1\leq i<j\leq n$, by \cite[Lem. 2.4(i)]{cheng:2014}, we have
\beao
&&\PP\left(
{\bf X}^{(i)}\Theta_i\in xA,\,\
{\bf X}^{(j)}\Theta_j\in xA
\right)
\\
&&=\PP\left(X_A^{(i)}\Theta_i>x,\,\
X_A^{(j)}\Theta_j>x
\right)
\\
&&=o\left[
\PP\left(X_A^{(i)}\Theta_i>x\right)+\PP\left(X_A^{(j)}\Theta_j>x\right)
\right]\\
&&=o\left[\PP\left({\bf X}^{(i)}\Theta_i\in xA\right)+\PP\left({\bf X}^{(j)}\Theta_j\in xA\right)\right].
\eeao
 Since $n$ is fixed, it follows that
\[
\sum_{1\leq i<j\leq n}
\PP\left(
{\bf X}^{(i)}\Theta_i\in xA,\,
{\bf X}^{(j)}\Theta_j\in xA
\right)=
o\left[
\sum_{i=1}^n
\PP\left({\bf X}^{(i)}\Theta_i\in xA\right)
\right].
\]
Therefore,
\[
\PP\left(\sum_{i=1}^n{\bf X}^{(i)}\Theta_i\in xA\right)
\gtrsim
\sum_{i=1}^n
\PP\left({\bf X}^{(i)}\Theta_i\in xA\right).
\]
Combining this with \eqref{eq.CKP.3.12} yields \eqref{eq.CKP.3.10}.

Next, we will prove \eqref{eq.CKP.3.11}. By \cite[Lem. 3.5]{li:tang:2015} (see, Remark \eqref{rem.CKP.3.2a}), for each $i=1,\ldots,n$, it holds that
\beam \label{eq.CKP.3.14}
\PP\left({\bf X}^{(i)}\Theta_i\in xA\right)
\sim
\bF_A(x)\E\left[\Theta_i^\alpha\right]=\PP\left({\bf X}^{(i)}\in xA\right)\E\left[\Theta_i^\alpha\right].
\eeam
Combining \eqref{eq.CKP.3.10} with \eqref{eq.CKP.3.14} yields
\eqref{eq.CKP.3.11}.
~\halmos

In the next lemma we shall establish relation \eqref{eq.CKP.3.11}, under the regular varying framework and with the more strict (moment) conditions on $\{\Theta_i,\,i\in\mathbb{N}\}$.

\ble \label{lem.CKP.3.1a}
Let $A\in\mathscr{R}$ be a fixed set, and let
$\{\Theta_i,\,i\in\bbn\}$ be a sequence of nonnegative and non-degenerate to zero random variables, independent of
$\{{\bf X}^{(i)},\,i\in\bbn\}$. Suppose that Assumption~\ref{ass.CKP.3.1} holds, $F\in\mathcal{R}_{-\alpha,A}$, for some $\alpha\geq 0$, and for some $q>\alpha$ it holds that
\beam \label{eq.CKP.3.7b}
\E[\Theta_i^{q}]<\infty,\quad i\in\bbn.
\eeam
Then relation \eqref{eq.CKP.3.11} holds for every fixed $n\in\mathbb{N}$.
\ele

\pr~
By the fact that $\mathcal{R}_{-\alpha,A}\subsetneq\mathcal{C}_A$, and \eqref{eq.CKP.3.7b} implies \eqref{eq.CKP.3.7a}, we obtain relation \eqref{eq.CKP.3.10} holds for every fixed $n\in\mathbb{N}$. Further, by \eqref{eq.CKP.3.7b} we can apply the Breiman's theorem (see, Remark \eqref{rem.CKP.3.2a}), to obtain relation \eqref{eq.CKP.3.14}, which in combination with \eqref{eq.CKP.3.10}, gives the desired result.
~\halmos

The third lemma shows that the contribution from the remaining infinite tail is asymptotically negligible.
\begin{lemma}\label{lem.CKP.3.2}
Under the conditions of Theorem~\ref{th.CKP.3.1a} (or, of Theorem~\ref{th.CKP.3.1}), for fixed $T\in\Lambda$, it holds that
\begin{align}\label{eq.CKP.3.115}
\lim_{M\to\infty}\limsup
\frac{\mathbb{P}\left(
\sum_{i=M+1}^{\infty}
X_A^{(i)}\Theta_i(T)>x
\right)}{
\PP\left({\bf X} \in xA\right)}=0.
\end{align}
Consequently,
\begin{align*}
\lim_{M\to\infty}\limsup
\frac{\mathbb{P}\left(\sum_{i=M+1}^{\infty}\mathbf X^{(i)}\Theta_i(T)\in xA\right)}{\PP\left({\bf X} \in xA\right)}=0.
\end{align*}
\end{lemma}

\pr~
We first prove our assertions under the conditions of Theorem \ref{th.CKP.3.1}. It suffices to prove \eqref{eq.CKP.3.115}, since for any $x\geq 0$ it holds that
\[
\PP\left(\sum_{i=M+1}^{\infty} {\bf X}^{(i)}\,\Theta_i(T) \in x\,A\right)
\leq \PP\left(\sum_{i=M+1}^{\infty} X_{A}^{(i)}\,\Theta_i(T) > x\right).
\]
Choose $b$, $\delta$, and $M$ such that
\(\alpha<b<p,\,1<\delta<{p}/{b}\),  and 
\(\sum_{i=M+1}^{\infty}i^{-\delta}\leq1
\), where $p$ is indicated in Assumption~\ref{ass.CKP.3.3}. 
Since the distribution of $X_A\Theta^*(T)$ belongs to $\mathcal{R}_{-\alpha}^*$ (by \cite[Lem. 3.5]{li:tang:2015}), Lemma 1 in \cite{li:2018} yields that, for some constant $C>0$,
\[
\PP\left(X_A^{(1)}\Theta^*(T)>i^{-\delta}x\right)
\leq
C i^{\delta b}
\PP\left(X_A^{(1)}\Theta^*(T)>x\right).
\]
Consequently,
\begin{align*}
& \PP\left(\sum_{i=M+1}^{\infty} X_{A}^{(i)}\,\Theta_i(T) > x\right) \leq \PP\left(\sum_{i=M+1}^{\infty} X_{A}^{(i)}\,\Theta_i(T) > \sum_{i=M+1}^{\infty} i^{-\delta}\,x\right) \\ \notag
&\leq \sum_{i=M+1}^{\infty} \PP\left(X_{A}^{(i)}\,\Theta_i(T) > i^{-\delta}\,x\right) \leq \sum_{i=M+1}^{\infty} \PP\left(X_{A}^{(i)}\,\Theta^*(T) > i^{-\delta}\,x\,,\; \tau_i \leq T\right) \\ \notag
&= \sum_{i=M+1}^{\infty} \PP\left( X_{A}\,\Theta^*(T) > i^{-\delta}\,x\right)\,\PP(\tau_i \leq T)  \\ \notag
&\leq C\,\PP\left( X_{A}\,\Theta^*(T) >x\right)\,\sum_{i=M+1}^{\infty}  i^{\delta\,b}\,\PP(\tau_i \leq T) \\ \notag
&=C\, \PP\left({\bf X}\,\Theta^*(T) \in xA\right)\,\sum_{i=M+1}^{\infty}  i^{\delta\,b}\,\PP(\tau_i \leq T),
\end{align*}  
Since $\delta b<p$, Assumption~\ref{ass.CKP.3.3} and Remark \ref{rem.CKP.3.2,5} imply that
\[
\sum_{i=M+1}^{\infty}
i^{\delta b}\PP(\tau_i\leq T)
\leq
\sum_{i=M+1}^{\infty}
i^p\PP(\tau_i\leq T)
\longrightarrow0
\]
as $M\to\infty$. Moreover, by \eqref{eq.CKP.Breiman}, it holds that
\begin{align}\label{eq.CKP.3.116}
\PP\left(X_A\Theta^*(T)>x\right)
\sim
\bF_A(x)\E\left[(\Theta^*(T))^\alpha\right]=\PP\left({\bf X} \in xA\right)\E\left[(\Theta^*(T))^\alpha\right].
\end{align}
Therefore, relation \eqref{eq.CKP.3.115} holds.

For the case of conditions of Theorem \eqref{th.CKP.3.1a}, relation \eqref{eq.CKP.3.115} follows with the same manner.  The only difference is that \eqref{eq.CKP.3.116} holds, and the distribution of $X_A\Theta^*(T)$ belongs to $\mathcal{R}_{-\alpha}$, by the using of Breiman's theorem.
~\halmos

The next lemma extends relation \eqref{eq.CKP.3.11} uniformly with respect to $n \in \bbn$, and plays a crucial role for the proofs of Theorems \ref{th.CKP.3.1a} and \ref{th.CKP.3.1}.

\ble \label{lem.CKP.3.3}
Under the conditions of Theorem~\ref{th.CKP.3.1a} (or, of Theorem~\ref{th.CKP.3.1}), for fixed $T\in\Lambda$, relation \eqref{eq.CKP.3.11} holds uniformly for $n\in\bbn$, namely,
\beam \label{eq.CKP.3.18}
\lim
\sup_{n\in\bbn}
\left|\dfrac{\PP\left(\sum_{i=1}^{n}{\bf X}^{(i)}\Theta_i(T)\in xA\right)}{\PP({\bf X}\in xA)\sum_{i=1}^{n}\E\left[\Theta_i^\alpha(T)\right]}-1\right|=0.
\eeam
\ele

\pr~
We first prove \eqref{eq.CKP.3.18}, under the conditions of Theorem \ref{th.CKP.3.1}. According to Remark~\ref{rem.CKP.3.4}, we have
$\sum_{i=1}^\infty\E\left[\Theta_i^\alpha(T)\right]<\infty$.
We first show that
\beam \label{eq.CKP.3.20}
\limsup
\frac{
\PP\left(\sum_{i=1}^\infty
{\bf X}^{(i)}\Theta_i(T)\in xA\right)}{\PP({\bf X}\in xA)}
\leq\sum_{i=1}^\infty\E\left[\Theta_i^\alpha(T)\right]
<\infty.
\eeam
For any $\varepsilon\in(0,1)$ and $M\in\bbn$,  by Lemma~\ref{lem.CKP.3.1}, we obtain 
\begin{align*}
&\limsup
\frac{
\PP\left(\sum_{i=1}^{\infty}{\bf X}^{(i)}\Theta_i(T)\in xA\right)}{\PP({\bf X}\in xA)}
\\
&\leq
\limsup
\frac{\PP\left(\sum_{i=1}^{\infty}X_A^{(i)}\Theta_i(T)>x,\,\sum_{i=M+1}^{\infty}X_A^{(i)}\Theta_i(T)\leq\varepsilon x\right)}{
	\bF_A(x)}
\\
& +\limsup\frac{\PP\left(\sum_{i=1}^{\infty}X_A^{(i)}\Theta_i(T)>x,\,\sum_{i=M+1}^{\infty}X_A^{(i)}\Theta_i(T)>\varepsilon x\right)}{\bF_A(x)}
\\
&\leq
\limsup
\frac{\PP\left(\sum_{i=1}^{M}X_A^{(i)}\Theta_i(T)>(1-\varepsilon)x\right)}{
\bF_A(x)}+\limsup\frac{\PP\left(\sum_{i=M+1}^{\infty}X_A^{(i)}\Theta_i(T)>\varepsilon x\right)}{\bF_A(x)}
\\
&=(1-\varepsilon)^{-\alpha}
\sum_{i=1}^{M}\E\left[\Theta_i^\alpha(T)\right]+\varepsilon^{-\alpha}\limsup\frac{\PP\left(\sum_{i=M+1}^{\infty}X_A^{(i)}\Theta_i(T)>\varepsilon x\right)}{\PP\left({\bf X} \in \varepsilon xA\right)}.
\end{align*}
According to Lemma~\ref{lem.CKP.3.2}, letting $M\to\infty$ and then
$\varepsilon\downarrow0$, we obtain \eqref{eq.CKP.3.20}.

Now, we fix some $N\in\bbn$. For every $n>N$, since all summands are nonnegative
and $xA$ is increasing,
\[
\PP\left(\sum_{i=1}^N{\bf X}^{(i)}\Theta_i(T)\in xA\right)\leq\PP\left(\sum_{i=1}^n{\bf X}^{(i)}\Theta_i(T)\in xA\right)\leq\PP\left(\sum_{i=1}^\infty{\bf X}^{(i)}\Theta_i(T)\in xA\right).
\]
Since
\[
\sum_{i=1}^N\E\left[\Theta_i^\alpha(T)\right]\leq\sum_{i=1}^n\E\left[\Theta_i^\alpha(T)\right]\leq\sum_{i=1}^\infty\E\left[\Theta_i^\alpha(T)\right],
\]
Lemma~\ref{lem.CKP.3.1} and \eqref{eq.CKP.3.20} imply
\begin{align*}
\frac{\sum_{i=1}^N\E\left[\Theta_i^\alpha(T)\right]}{\sum_{i=1}^\infty\E\left[\Theta_i^\alpha(T)\right]}
&\leq
\liminf\inf_{n>N}
\frac{\PP\left(\sum_{i=1}^n{\bf X}^{(i)}\Theta_i(T)\in xA\right)}{\PP({\bf X}\in xA)\sum_{i=1}^n\E\left[\Theta_i^\alpha(T)\right]}
\\
&\leq
\limsup\sup_{n>N}
\frac{\PP\left(\sum_{i=1}^n{\bf X}^{(i)}\Theta_i(T)\in xA\right)}{\PP({\bf X}\in xA)\sum_{i=1}^n\E\left[\Theta_i^\alpha(T)\right]}
\leq\frac{\sum_{i=1}^\infty\E\left[\Theta_i^\alpha(T)\right]}{\sum_{i=1}^N\E\left[\Theta_i^\alpha(T)\right]}.
\end{align*}
Since
\[
\sum_{i=1}^N\E\left[\Theta_i^\alpha(T)\right]
\uparrow
\sum_{i=1}^\infty\E\left[\Theta_i^\alpha(T)\right],
\]
letting $N\to\infty$ gives
\[
\lim_{N\to\infty}
\limsup
\sup_{n>N}\left|\frac{\PP\left(\sum_{i=1}^n{\bf X}^{(i)}\Theta_i(T)\in xA\right)}{
\PP({\bf X}\in xA)\sum_{i=1}^n\E\left[\Theta_i^\alpha(T)\right]}-1\right|=0.
\]
On the other hand, for every fixed $N$, Lemma~\ref{lem.CKP.3.1} yields
\[
\lim
\max_{1\leq n\leq N}
\left|\frac{\PP\left(\sum_{i=1}^n{\bf X}^{(i)}\Theta_i(T)\in xA\right)}{\PP({\bf X}\in xA)\sum_{i=1}^n\E\left[\Theta_i^\alpha(T)\right]}-1\right|=0.
\]
Combining the last two relations proves \eqref{eq.CKP.3.18}.

Under the conditions of Theorem \ref{th.CKP.3.1a}, \eqref{eq.CKP.3.18} can be derived by similar methodology, with the only difference the using of Lemma \eqref{lem.CKP.3.1a} instead of Lemma \eqref{lem.CKP.3.1}.
~\halmos

Now, we prove our main results.

\noindent{\bf Proof of Theorem~\ref{th.CKP.3.1a}.}~
Recall that
\[
{\bf D}(T)=
\sum_{i=1}^{\infty}{\bf X}^{(i)}\Theta_i(T).
\]
By Lemma~\ref{lem.CKP.3.3} together with
$\sum_{i=1}^{\infty}\E[\Theta_i^\alpha(T)]<\infty$ from
Remark~\ref{rem.CKP.3.4}, we obtain by letting $n\to\infty$ that
\[
\PP\left({\bf D}(T)\in xA\right)
\sim
\PP({\bf X}\in xA)\sum_{i=1}^{\infty}\E\left[\Theta_i^\alpha(T)\right].
\]
Moreover, by the mutual independence of the counting process and the
investment return process,
\begin{align*}
\sum_{i=1}^{\infty}\E\left[\Theta_i^\alpha(T)\right]
&=\sum_{i=1}^{\infty}
\int_{0}^T
\E\left[e^{-\alpha\xi(s)}\right]\PP(\tau_i\in ds)
\\
&=
\int_0^T
\E\left[e^{-\alpha\xi(s)}\right]\lambda(ds).
\end{align*}
Therefore,
\[
\PP\left({\bf D}(T)\in xA\right)
\sim
\PP({\bf X}\in xA)
\int_0^T
\E\left[e^{-\alpha\xi(s)}\right]\lambda(ds),
\]
which proves \eqref{eq.CPK.main}.
~\halmos

\noindent{\bf Proof of Theorem~\ref{th.CKP.3.1}.}~
The proof is the same as that of Theorem \ref{th.CKP.3.1a}.
~\halmos

\section{Numerical study}
\label{sec.CKP.4}
In this section, we conduct a numerical study to illustrate the finite-time
asymptotic relation established in Theorem~\ref{th.CKP.3.1}. In particular,
we consider a bivariate claim-size distribution that satisfies $\mathcal{R}_{-\alpha,A}^*$ on a fixed set $A\in\mathscr{R}$, but does not belong to the classical $\operatorname{MRV*}$. We first
specify the claim-size distribution, the claim-arrival, and stochastic investment return processes, and
then describe the Monte Carlo procedure. The baseline experiment is followed
by a sensitivity analysis with respect to the main model parameters.

\subsection{Model specification}
Let
\(A=\left\{{\bf x}\in\bbr^2:x_1\vee x_2>1\right\}\in\mathscr R,\)
and let $\{{\bf Y}^{(i)},\,i\in\bbn\}$ be independent copies of the random
vector constructed in Example~\ref{eg.CKP.eg1}. More precisely, let
$\{Z_i,\,i\in\bbn\}$ be i.i.d. positive random variables with survival
function
\[
\bH(x)=x^{-\alpha}(1+\log x)^{-\beta},\qquad x\geq1,
\]
where $\alpha>0$ and $\beta>1$. Then $H\in\mathcal R^{\ast}_{-\alpha}$. Define
\[
{\bf Y}^{(i)}=
\begin{cases}
(Z_i,0), & c^{2k}\leq Z_i<c^{2k+1},\\
(0,Z_i), & c^{2k+1}\leq Z_i<c^{2k+2},
\end{cases}
\qquad k=0,1,\ldots,
\]
for some $c>1$.  Furthermore, let ${\bf W}=(W_1,W_2)$, where $W_1$ and
$W_2$ are independent $U(0,1)$ random variables, and assume that
${\bf W}$ is independent of $\{{\bf Y}^{(i)},\,i\in\bbn\}$. The claim
vectors are defined by
\[
{\bf X}^{(i)}={\bf Y}^{(i)}+{\bf W},\qquad i\in\bbn,
\]
where the same random vector ${\bf W}$ is shared by all claims. 
Hence, $\{{\bf X}^{(i)},\,i\in\bbn\}$ are identically distributed but dependent. Denote the common distribution of $\{{\bf X}^{(i)},\,i\in\bbn\}$ by $F$. We obtain from the argument in Example~\ref{eg.CKP.eg1} that $F$ is not $\operatorname{MRV}$. Since
$0\leq W_1\vee W_2\leq1$ almost surely, we have
\[
Z_i\leq X_A^{(i)}\leq Z_i+1,\,\,\, i\in\mathbb{N},
\]
almost surely. Therefore,
\[
\bH(x)\leq\bF_A(x)\leq\bH(x-1),
\]
and, since $H\in\mathcal R^{\ast}_{-\alpha}\subseteq \mathcal L$,
\[
\bF_A(x)\sim\bH(x).
\]
Hence, $F_A\in\mathcal{R}_{-\alpha}^*$. Moreover, in view of
Example~\ref{eg.CKP.eg2}, the claim sequence satisfies the $TQAI_A$
condition.

For the claim-number process, we consider a homogeneous Poisson process
$\{N(t),\,t\geq0\}$ with intensity $\lambda>0$. Therefore, its mean
measure is given by
\(
\lambda(\dd s)=\lambda\,\dd s.
\)

For the stochastic investment return process, we consider the
jump-diffusion L\'evy process
\[
\xi(t)=
\mu t+\sigma W(t)
-\sum_{j=1}^{N_J(t)}J_j,
\qquad t\geq0,
\]
where, $\mu\in\mathbb{R}$, $\sigma>0$, $\{W(t),\,t\geq0\}$ is a Wiener process,
$\{N_J(t),\,t\geq0\}$ is a Poisson process with intensity
$\lambda_J>0$, and $\{J_j,\,j\in\mathbb{N}\}$ are i.i.d. exponential random
variables with parameter $\beta_J>\alpha$. The claim sequence
$\{{\bf X}^{(i)},\,i\in\bbn\}$, the claim-arrival process
$\{N(t),\,t\geq0\}$, and the investment return process
$\{\xi(t),\,t\geq0\}$ are mutually independent. In addition,
$W$, $N_J$, and $\{J_j\}$ are mutually independent. Then, the Laplace
exponent of $\{\xi(t),\,t\geq0\}$ satisfies for every $\alpha\leq q<\beta_J$,
\[
\phi(q)=
-q\mu+\frac{q^2\sigma^2}{2}
+\lambda_J
\left(
\frac{\beta_J}{\beta_J-q}-1
\right)<\infty.
\]
By the last relation, and Remark \ref{rem.CKP.3.2}, the moment condition of Assumption \ref{ass.CKP.3.2a} is true (so relation \eqref{eq.CKP.3.3} is also valid), for the process $\xi$. By Theorem~\ref{th.CKP.3.1}, see also Remark \ref{rem.CKP.3.5}, and the tail equivalence $\bF_A(x)\sim\bH(x)$, we therefore have
\begin{align} \label{eq.CKP.sim.5}
&\quad \PP({\bf D}(T)\in xA)
\sim\bH(x)\lambda
\int_0^T \e^{s\phi(\alpha)}\,\dd s\notag\\[2mm]
&=:\operatorname{APP}(x;T)=
\begin{cases}
\displaystyle
\lambda\frac{\e^{T\phi(\alpha)}-1}{\phi(\alpha)}\,\bH(x),
& \phi(\alpha)\neq0,\\[2mm]
\lambda T\, \bH(x),
& \phi(\alpha)=0.
\end{cases}
\end{align}
\subsection{Monte Carlo procedure}

For each simulated sample path, we first generate
\[
N(T)\overset{d}{\sim}\operatorname{Poisson}(\lambda T).
\]
Conditional on $N(T)=n$, the claim-arrival times are generated as the order
statistics of $n$ independent uniform random variables on $[0,T]$, see for example \cite[Th. 5.2]{ross:2014}.

The L\'evy process $\xi$ is simulated exactly at the claim-arrival times.
For two consecutive arrival times with time increment $\Delta$, the
independent and stationary increment property gives
\beao
\Delta\xi=
\mu\Delta+\sigma\sqrt{\Delta}\,\eta
-\sum_{j=1}^{M_\Delta}J_j,
\eeao
where $\eta\overset{d}{\sim} N(0,1)$,
$M_\Delta\overset{d}{\sim}\operatorname{Poisson}(\lambda_J\Delta)$, and
$\{J_j\}$ are i.i.d. exponential random variables with parameter
$\beta_J$. Moreover, they are mutually independent. 

For the $j$th Monte Carlo sample path, the discounted aggregate claim vector
is calculated as
\beao
{\bf D}_j(T)=
\sum_{i=1}^{N_j(T)}
{\bf X}_j^{(i)}\mathrm{e}^{-\xi_j(\tau_i)}.
\eeao
Since
\beao
{\bf D}_j(T)\in xA
\quad\Longleftrightarrow\quad
D_{1,j}(T)\vee D_{2,j}(T)>x,
\eeao
the required probability can be estimated directly by empirical frequencies.

For each parameter setting, the Monte Carlo experiment is independently
repeated $N_{\rm rep}=50$ times. In each replication, $m=100000$ independent sample
paths are generated. For the $r$th replication, define
\[
\widehat P_r(x)=\frac{1}{m}
\sum_{j=1}^{m}
{\bf 1}_{\{{\bf D}_{r,j}(T)\in xA\}},
\qquad r=1,\ldots,N_{\rm rep}.
\]
The final Monte Carlo estimator is
\[
\overline P(x)=
\frac{1}{N_{\rm rep}}
\sum_{r=1}^{N_{\rm rep}}
\widehat P_r(x).
\]
For each replication, the same simulated sample is used for all threshold
levels. We also define
\[
R_r(x)=
\frac{\widehat P_r(x)}
{\operatorname{APP}(x;T)},
\qquad
\overline R(x)=
\frac{1}{N_{\rm rep}}
\sum_{r=1}^{N_{\rm rep}}R_r(x).
\]
According to Theorem~\ref{th.CKP.3.1}, $\overline R(x)$ is expected to
approach one as $x$ increases.

\subsection{Baseline case}

For the baseline experiment, we take
\beao
\alpha=2,\qquad
\beta=1.2,\qquad
c=2,\qquad
\lambda=1,\qquad
T=1,
\eeao
and
\beao
\mu=0.05,\qquad
\sigma=0.20,\qquad
\lambda_J=0.50,\qquad
\beta_J=5.
\eeao

The threshold levels are chosen as high quantiles of $H$:
\beao
q\in
\{0.99,0.995,0.997,0.998,0.999,0.9995,0.9998\},
\qquad
x_q=H^{-1}(q).
\eeao
For each $x_q$, we compare $\overline P(x_q)$
with $\operatorname{APP}(x_q;T)$ in
\eqref{eq.CKP.sim.5}. The following figures present the numerical results for the baseline case.

\begin{figure}[htbp]
\centering
\begin{minipage}{0.48\textwidth}
\centering
\includegraphics[width=\textwidth]
{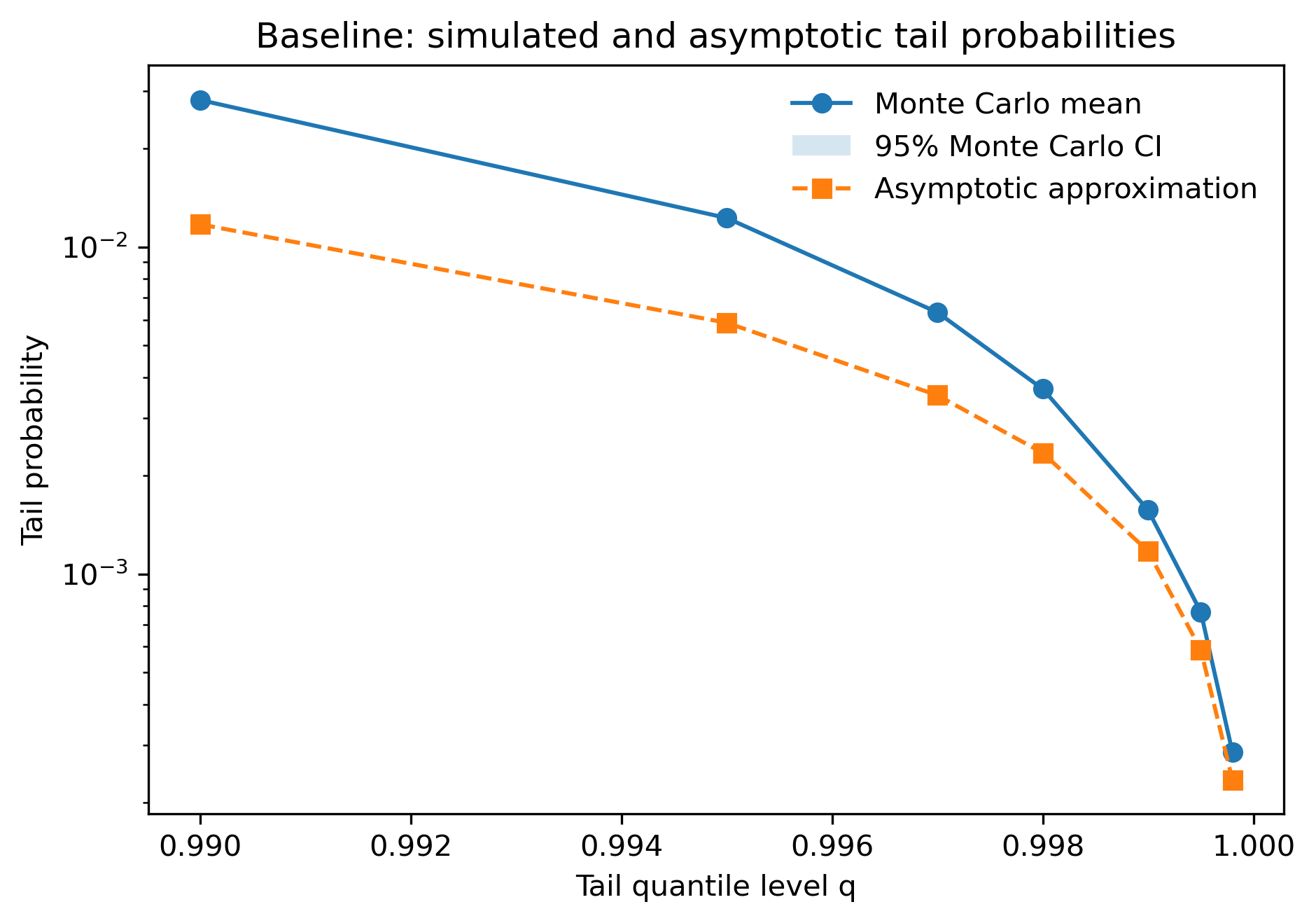}
\end{minipage}
\hfill
\begin{minipage}{0.48\textwidth}
\centering
\includegraphics[width=\textwidth]
{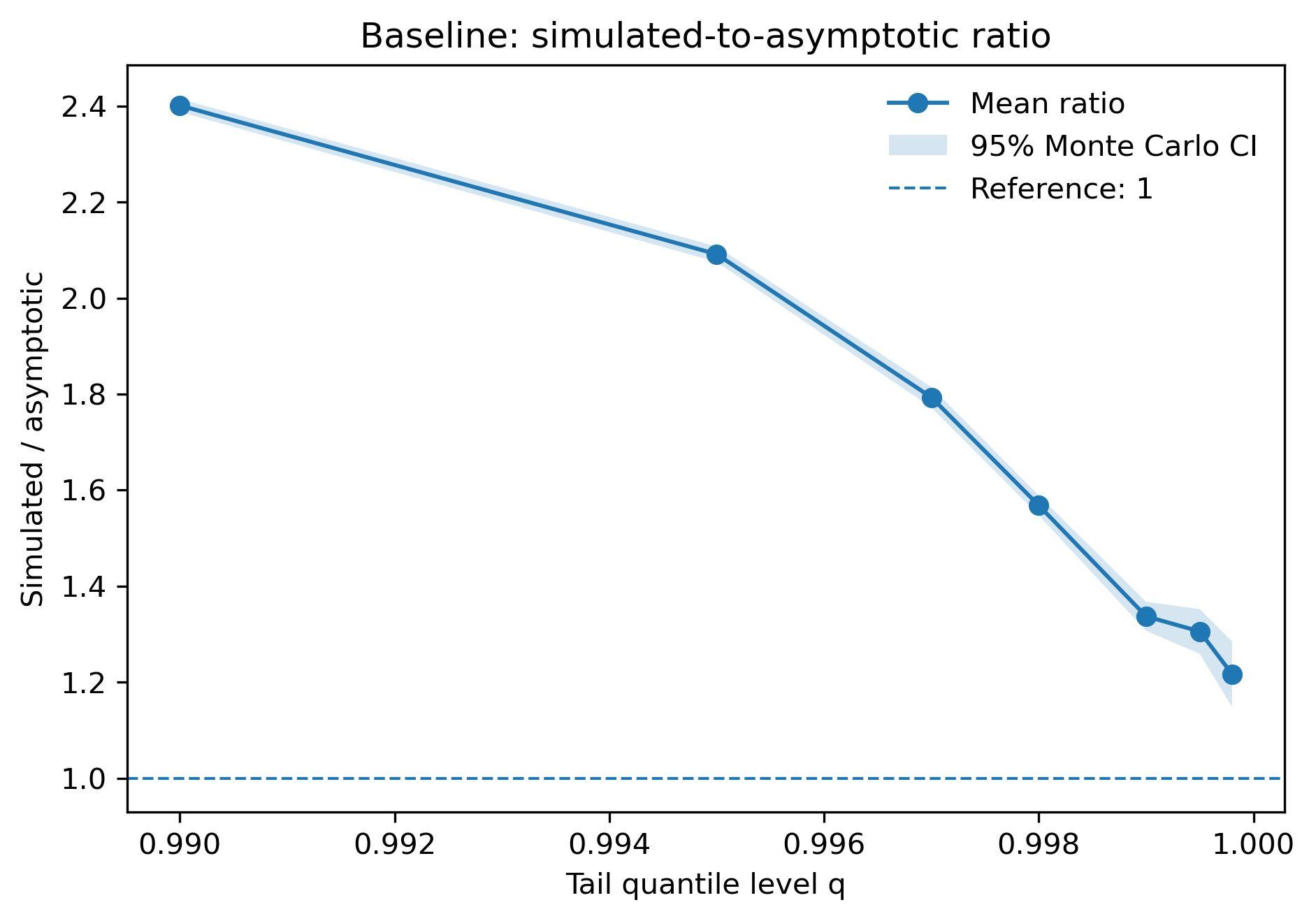}
\end{minipage}
\caption{Baseline numerical results.}
\label{fig.CKP.baseline}
\end{figure}
The pointwise 95\% Monte Carlo confidence intervals are constructed from the
$N_{\rm rep}$ replication estimates using
\[
\overline P(x)
\pm
1.96\frac{s_P(x)}{\sqrt{N_{\rm rep}}},
\]
where $s_P(x)$ denotes the sample standard deviation of
$\{\widehat P_r(x)\}_{r=1}^{N_{\rm rep}}$.

Figure~\ref{fig.CKP.baseline} shows that both the simulated and asymptotic probabilities decrease toward zero as $q$ increases. Moreover, their ratio approaches one as $q$ increases, which means the formula \eqref{eq.CKP.sim.5} provides a well-behaved estimate.
% After obtaining the final numerical output, summarize here the observed
% convergence pattern and the approximation accuracy at the highest
% quantile levels.

\subsection{Sensitivity analysis}

We next investigate the sensitivity of the numerical results to changes in
the model parameters. The graphical analysis focuses on three parameters:
the claim tail index $\alpha$, the jump intensity $\lambda_J$, and the
direction-switching parameter $c$. Specifically, we consider
\beao
\alpha\in\{1.8,2.0,2.2\},\qquad
\lambda_J\in\{0,0.5,1.0\},\qquad
c\in\{1.5,2.0,3.0\},
\eeao
while all remaining parameters are kept at their baseline values. The parameter $\alpha$ controls the heaviness of
the claim-size tail, while $\lambda_J$ determines the frequency of downward
jumps in the investment return process. The parameter $c$ changes the
alternating directional structure of the claim vector.

\begin{figure}[!htbp]
\centering

\begin{subfigure}[t]{0.48\textwidth}
\centering
\includegraphics[width=\textwidth]
{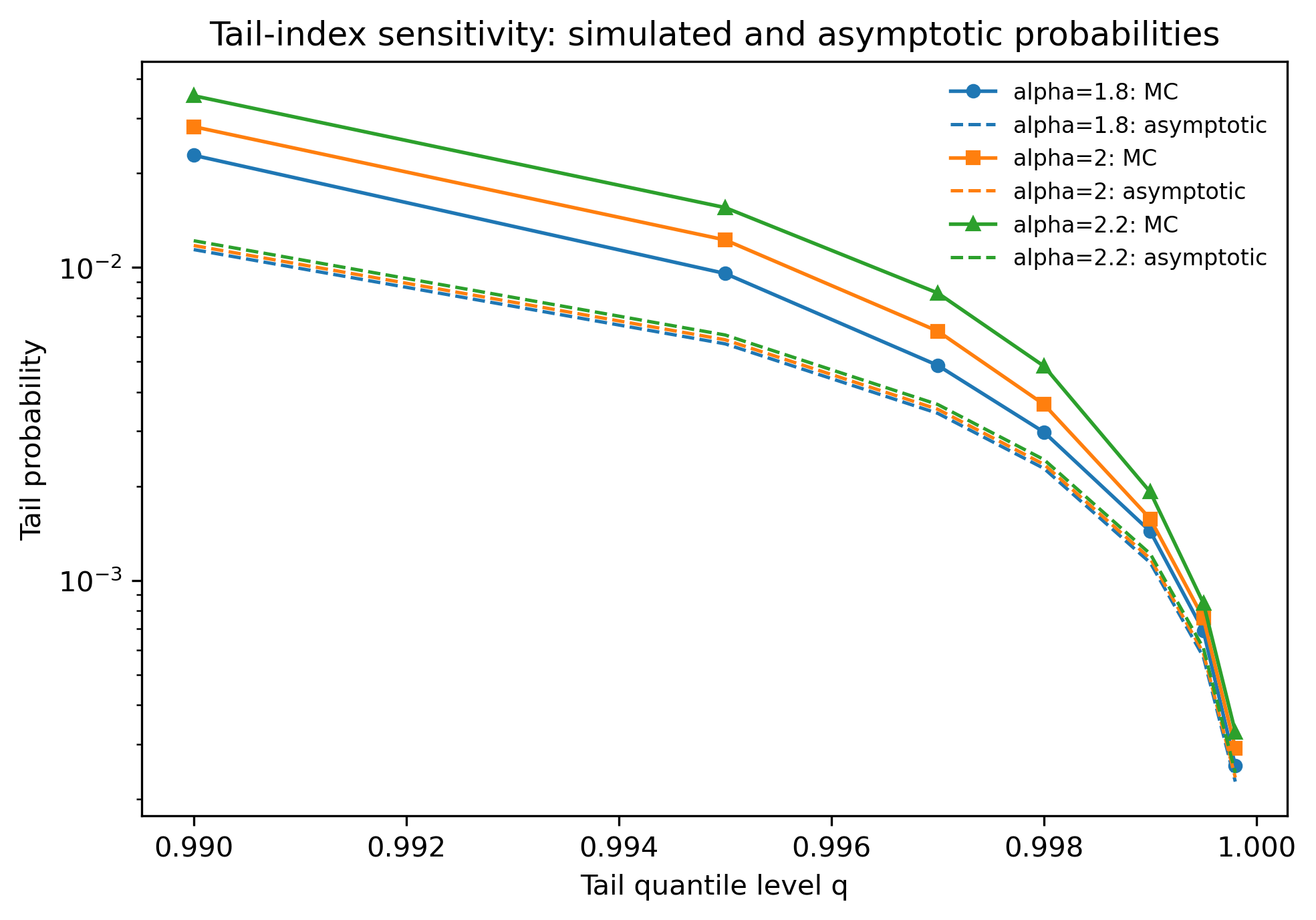}
\caption{Simulated and asymptotic probabilities for different $\alpha$.}
\label{fig.CKP.sen.alpha.prob}
\end{subfigure}
\hfill
\begin{subfigure}[t]{0.48\textwidth}
\centering
\includegraphics[width=\textwidth]
{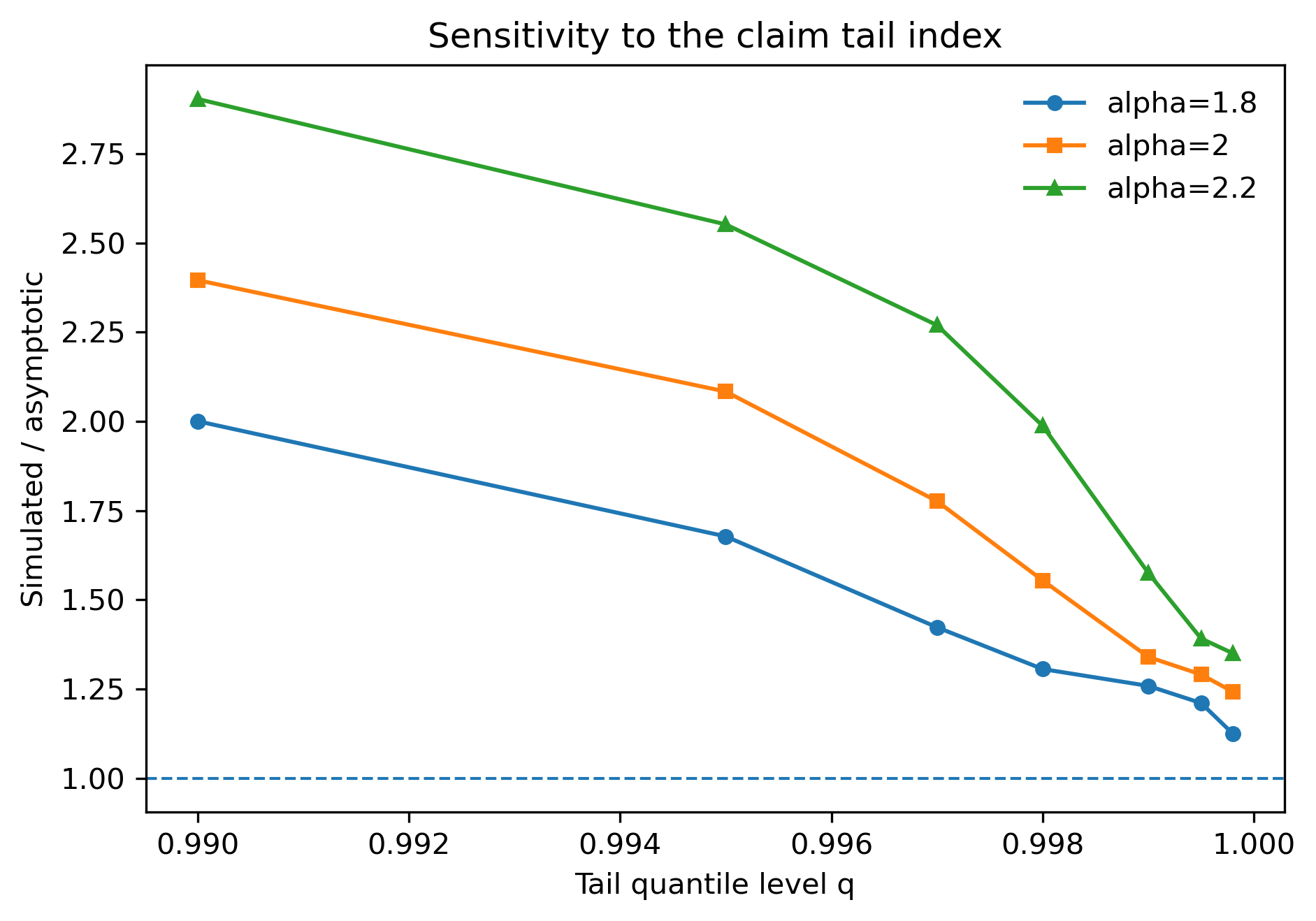}
\caption{Simulated-to-asymptotic ratios for different $\alpha$.}
\label{fig.CKP.sen.alpha.ratio}
\end{subfigure}

\vspace{2mm}

\begin{subfigure}[t]{0.48\textwidth}
\centering
\includegraphics[width=\textwidth]
{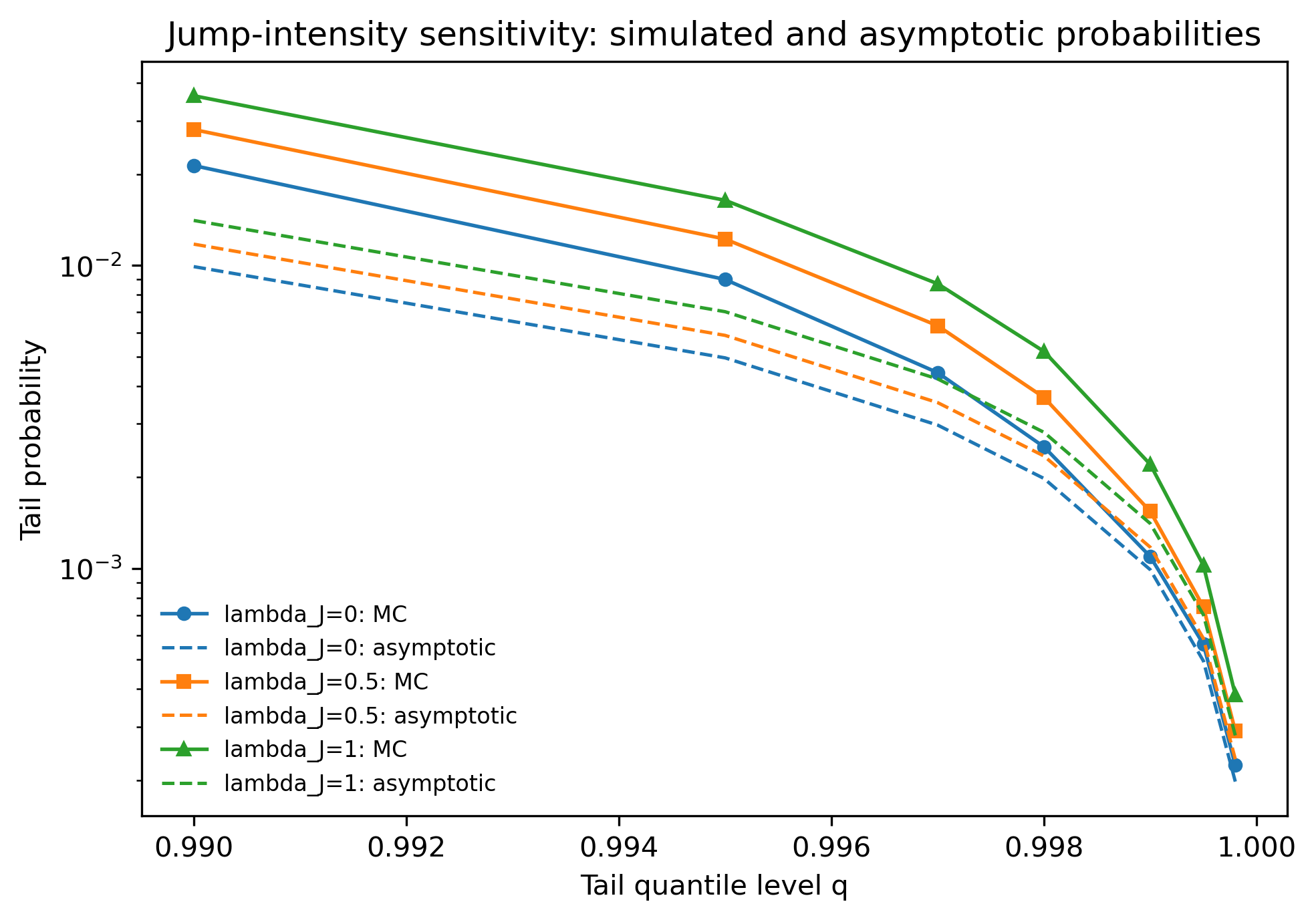}
\caption{Simulated and asymptotic probabilities for different $\lambda_J$.}
\label{fig.CKP.sen.lambdaJ.prob}
\end{subfigure}
\hfill
\begin{subfigure}[t]{0.48\textwidth}
\centering
\includegraphics[width=\textwidth]
{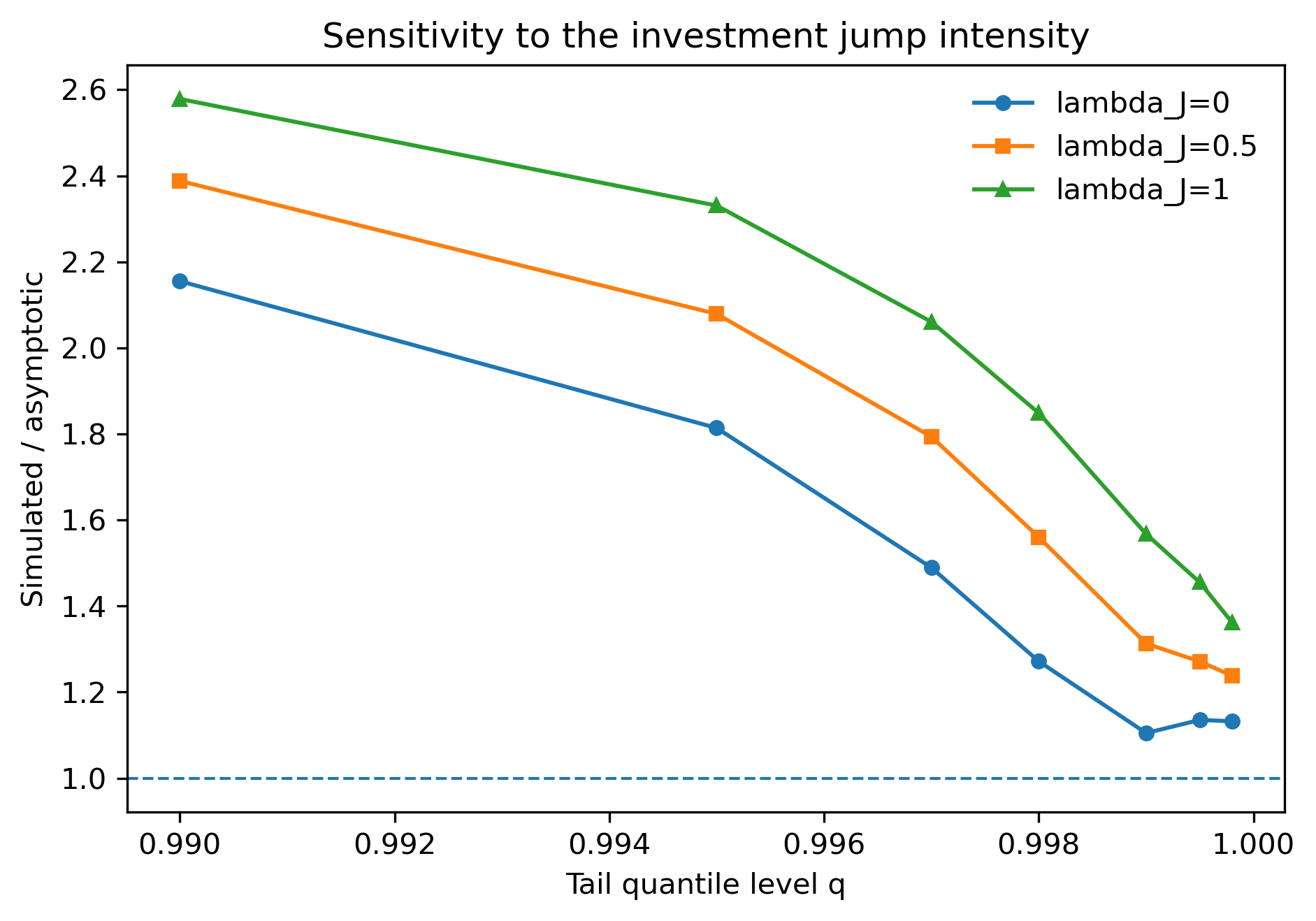}
\caption{Simulated-to-asymptotic ratios for different $\lambda_J$.}
\label{fig.CKP.sen.lambdaJ.ratio}
\end{subfigure}

\vspace{2mm}

\begin{subfigure}[t]{0.48\textwidth}
\centering
\includegraphics[width=\textwidth]
{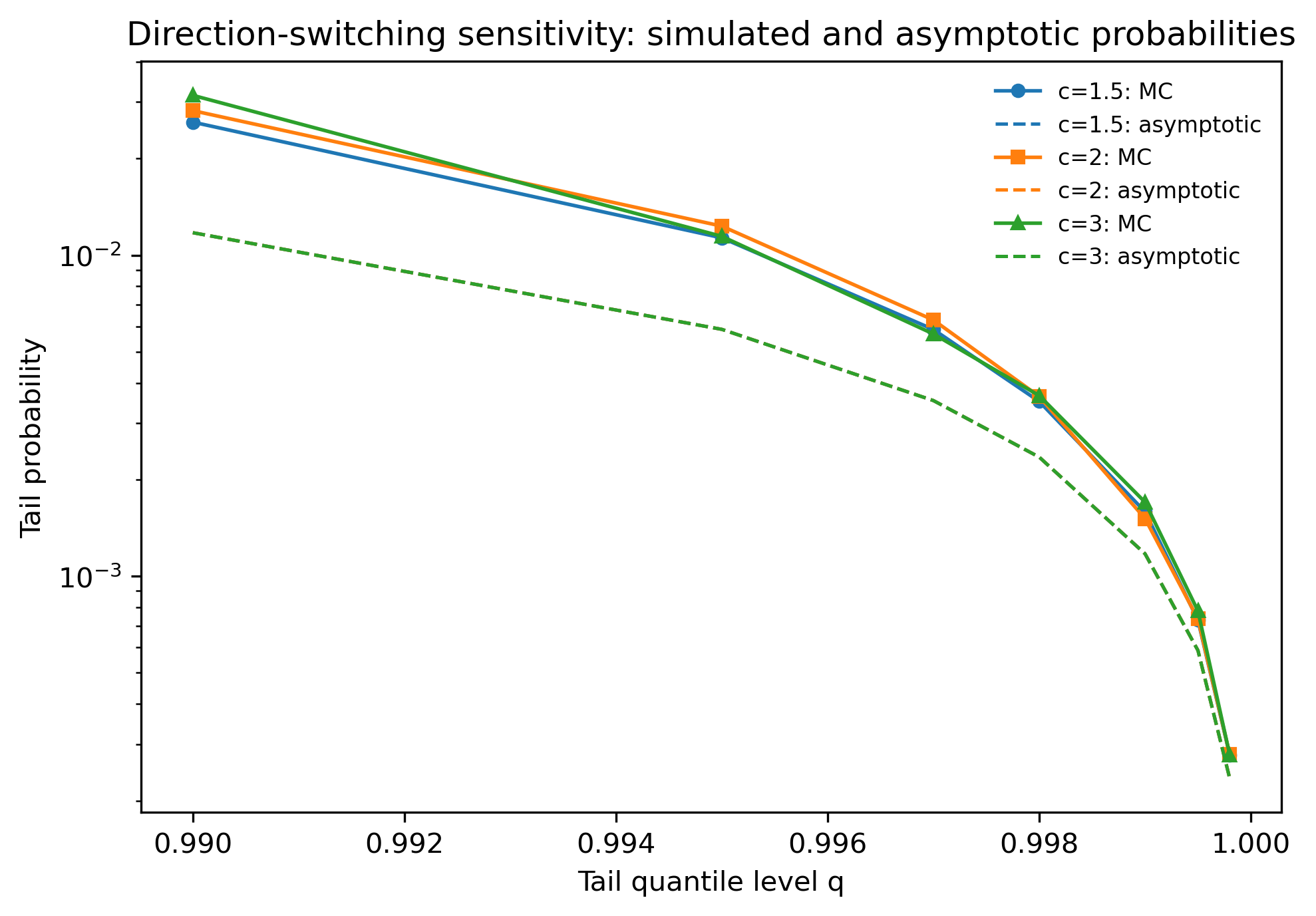}
\caption{Simulated and asymptotic probabilities for different $c$.}
\label{fig.CKP.sen.c.prob}
\end{subfigure}
\hfill
\begin{subfigure}[t]{0.48\textwidth}
\centering
\includegraphics[width=\textwidth]
{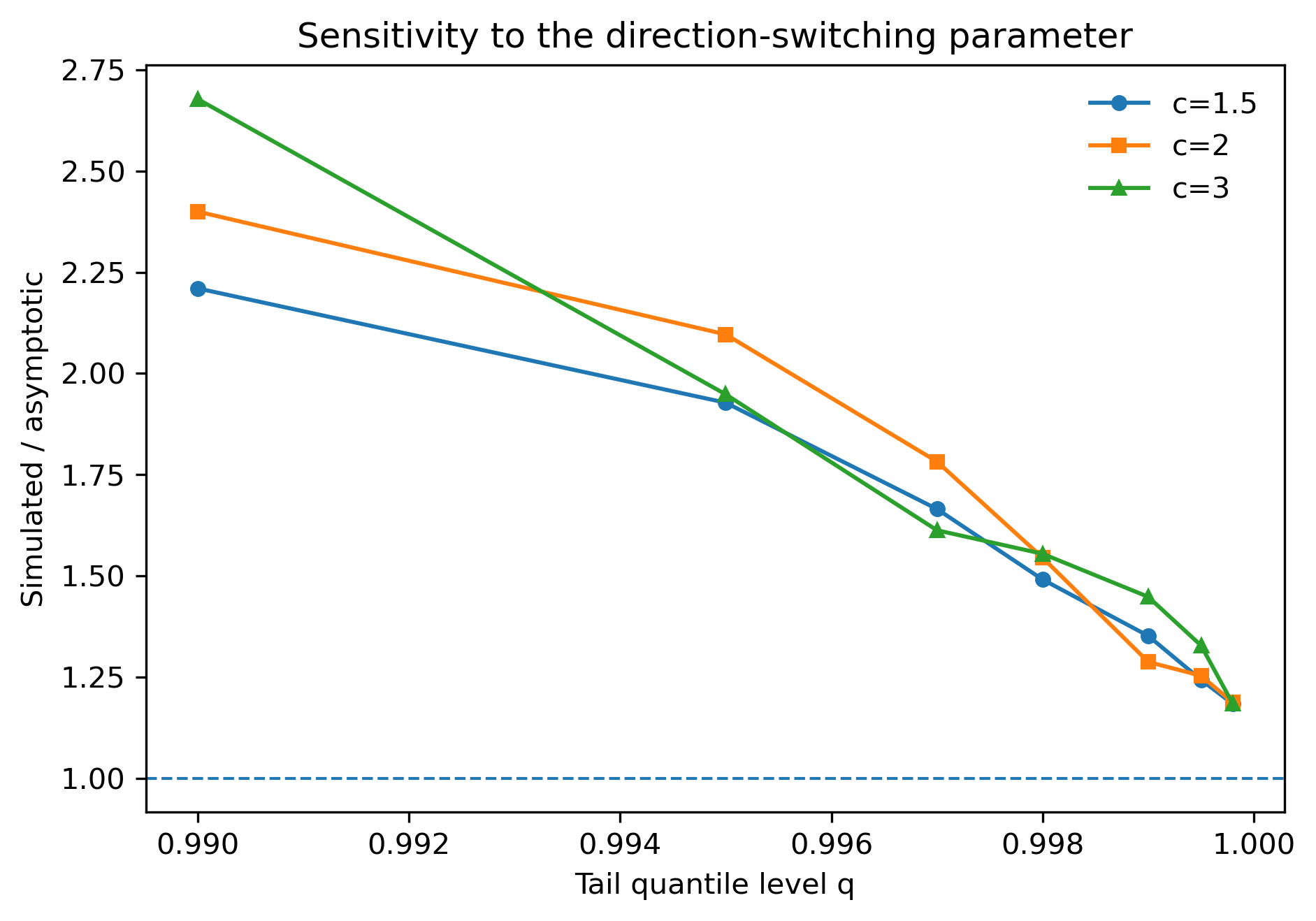}
\caption{Simulated-to-asymptotic ratios for different $c$.}
\label{fig.CKP.sen.c.ratio}
\end{subfigure}

\caption{Sensitivity analysis for the tail index $\alpha$, the jump intensity
$\lambda_J$, and the direction-switching parameter $c$.}
\label{fig.CKP.sensitivity}
\end{figure}

Figure~\ref{fig.CKP.sensitivity} reports the results of the sensitivity analysis.
For comparability across different parameter settings, the threshold levels
$x_q$ are kept fixed at those determined under the baseline model. 

First, the tail index $\alpha$ has a pronounced effect on the tail probability: a smaller $\alpha$ corresponds to a heavier claim-size tail and hence leads to a substantially larger tail probability. For all three values of $\alpha$, the ratios of simulated and asymptotic values decrease toward one as $q$ increases, although the convergence is somewhat slower for larger $\alpha$.

Second, increasing the jump intensity $\lambda_J$ increases the tail probability. This is consistent with the fact that more frequent downward jumps in $\xi$ generate larger discount factors $\e^{-\xi(t)}$, thereby increasing the probability of a large discounted aggregate claim. The corresponding ratios again move toward one in the high-quantile region, supporting the asymptotic formula in Theorem~\ref{th.CKP.3.1}. 

Finally, the parameter $c$ does not affect the first-order asymptotic approximation,
because the tail equivalence $\overline F_A(x)\sim\overline H(x)$ is independent of $c$.
Nevertheless, different values of $c$ alter the allocation of individual claims between
the two coordinates and may therefore produce visible finite-threshold differences in the
simulated probabilities. These differences seem to diminish at higher quantile levels, illustrating that the effect of $c$ is of lower order from the viewpoint of the set-based tail asymptotics.

In addition to the graphical analysis, we conduct a sensitivity analysis at a fixed high threshold. We set
\beao
x_0=H^{-1}(0.999).
\eeao
Keeping $x_0$
fixed, each model parameter is increased and decreased by $10\%$, while
all remaining parameters are kept at their baseline values. The same set
of $50$ Monte Carlo seeds is used across the sensitivity scenarios to
reduce the effect of simulation noise.

The resulting percentage changes are summarized in Table~\ref{tab.CKP.sensitivity}.

\begin{table}[htbp]
\centering
\caption{Sensitivity analysis at $q_0=0.999$.}
\label{tab.CKP.sensitivity}
\renewcommand{\arraystretch}{1.15}
\footnotesize
\begin{tabular}{ccccccc}
\toprule
Parameter
& Change
& Value
& Simulation
& $\Delta_{\rm MC}$ (\%)
& Asymptotic
& $\Delta_{\rm APP}$ (\%) \\
\cmidrule(lr){4-4}\cmidrule(lr){6-6}
&
&
&
$\times 10^{-3}$
&
&
$\times 10^{-3}$
& \\
\midrule

\multirow{3}{*}{$\alpha$}
& $+10\%$ & $2.20$ & $0.9828$ & $-36.07$ & $0.7126$ & $-39.32$ \\
& $0\%$   & $2.00$ & $1.5372$ & $0$      & $1.1744$ & $0$      \\
& $-10\%$ & $1.80$ & $2.4486$ & $59.29$  & $1.9453$ & $65.64$  \\
\midrule

\multirow{3}{*}{$\beta$}
& $+10\%$ & $1.32$ & $1.3362$ & $-13.08$ & $1.0046$ & $-14.46$ \\
& $0\%$   & $1.20$ & $1.5372$ & $0$      & $1.1744$ & $0$      \\
& $-10\%$ & $1.08$ & $1.7836$ & $16.03$  & $1.3728$ & $16.90$  \\
\midrule

\multirow{3}{*}{$\lambda$}
& $+10\%$ & $1.10$ & $1.7506$ & $13.88$  & $1.2918$ & $10.00$  \\
& $0\%$   & $1.00$ & $1.5372$ & $0$      & $1.1744$ & $0$      \\
& $-10\%$ & $0.90$ & $1.3466$ & $-12.40$ & $1.0570$ & $-10.00$ \\
\midrule

\multirow{3}{*}{$\mu$}
& $+10\%$ & $0.055$ & $1.5276$ & $-0.62$ & $1.1682$ & $-0.52$ \\
& $0\%$   & $0.050$ & $1.5372$ & $0$     & $1.1744$ & $0$     \\
& $-10\%$ & $0.045$ & $1.5492$ & $0.78$  & $1.1806$ & $0.53$  \\
\midrule

\multirow{3}{*}{$\sigma$}
& $+10\%$ & $0.22$ & $1.5706$ & $2.17$  & $1.1848$ & $0.89$  \\
& $0\%$   & $0.20$ & $1.5372$ & $0$     & $1.1744$ & $0$     \\
& $-10\%$ & $0.18$ & $1.5112$ & $-1.69$ & $1.1651$ & $-0.80$ \\
\midrule

\multirow{3}{*}{$\lambda_J$}
& $+10\%$ & $0.55$ & $1.6294$ & $6.00$  & $1.1952$ & $1.77$  \\
& $0\%$   & $0.50$ & $1.5372$ & $0$     & $1.1744$ & $0$     \\
& $-10\%$ & $0.45$ & $1.4944$ & $-2.78$ & $1.1540$ & $-1.73$ \\
\midrule

\multirow{3}{*}{$\beta_J$}
& $+10\%$ & $5.50$ & $1.4388$ & $-6.40$ & $1.1455$ & $-2.46$ \\
& $0\%$   & $5.00$ & $1.5372$ & $0$     & $1.1744$ & $0$     \\
& $-10\%$ & $4.50$ & $1.6900$ & $9.94$  & $1.2165$ & $3.59$  \\
\midrule

\multirow{3}{*}{$c$}
& $+10\%$ & $2.20$ & $1.5832$ & $2.99$ & $1.1744$ & $0$ \\
& $0\%$   & $2.00$ & $1.5372$ & $0$    & $1.1744$ & $0$ \\
& $-10\%$ & $1.80$ & $1.6300$ & $6.04$ & $1.1744$ & $0$ \\
\bottomrule
\end{tabular}
\end{table}
In Table~\ref{tab.CKP.sensitivity}, for a generic parameter $\vartheta$, the relative change in the simulated
tail probability is defined by
\beao
\Delta_{\rm MC}(\vartheta)=
\frac{\overline P_{\vartheta}(x_0)-
\overline P_{0}(x_0)}{\overline P_{0}(x_0)}
\times100\%,
\eeao
and the corresponding relative change in the asymptotic approximation is
defined by
\beao
\Delta_{\rm APP}(\vartheta)=\frac{\operatorname{APP}_{\vartheta}(x_0;T)-\operatorname{APP}_{0}(x_0;T)}{\operatorname{APP}_{0}(x_0;T)}
\times100\%.
\eeao
Table~\ref{tab.CKP.sensitivity} provides a quantitative summary of the sensitivity analysis at the fixed high threshold. The tail
probability is most sensitive to the tail index $\alpha$, followed by the
tail-shape parameter $\beta$ and the claim-arrival intensity $\lambda$.
The effects of the investment parameters $\mu$, $\sigma$, $\lambda_J$, and
$\beta_J$ are comparatively more moderate. In most cases, the simulated
and asymptotic probabilities respond in the same direction to parameter
changes, which further supports the qualitative accuracy of the
asymptotic approximation. In particular, variations in $c$ leave the
first-order asymptotic estimates unchanged, while the corresponding
simulated probabilities still exhibit finite-threshold fluctuations,
consistent with the observations in Figure~\ref{fig.CKP.sensitivity}.

\section*{Acknowledgments}
We feel the pleasant duty to express our sincere gratitude to Prof. Dimitrios G. Konstantinides, for several discussions, which substantially improved the paper.

\FloatBarrier
%\printbibliography

\end{document}